\documentclass[12pt]{article}
\usepackage{epsfig}
      \def\di{\displaystyle}
      
      \def\bp{{\bf p}}
      \def\br{{\bf r}}
      \def\bs{{\bf s}}

      \def\E{{\cal E}}
      \def\H{{\cal H}}

      \def\R{{\cal R}}

\begin{document}

\vskip 5mm
\begin{center}
{\large\bf NUCLEAR SCISSORS MODE WITH PAIRING }\\
\vspace*{1cm}
{\large E.B. Balbutsev, L.A. Malov}\\
\vspace*{0.2cm}
{\it Joint Institute for Nuclear Research, 141980 Dubna, Moscow Region,
Russia}\\
\vspace*{0.5cm}
{\large P. Schuck, M. Urban, }\\
\vspace*{0.2cm}
{\it Institut de Physique Nucleaire, CNRS and Univ. Paris-Sud, 91406 Orsay
  Cedex, France}\\
\vspace*{0.5cm}
{\large X. Vi\~nas }\\
\vspace*{0.2cm}
{\it Departament d'Estructura i Constituents de la Mat\'eria
Facultat de F\'{\i}sica,\\ Universitat de Barcelona Diagonal 647,
  08028 Barcelona, Spain}
\end{center}

      \vspace{3cm}

\begin{abstract}
The coupled dynamics of the scissors mode and the isovector giant
quadrupole resonance are studied using a generalized Wigner function
moments method taking into account pair correlations. Equations
of motion for angular momentum, quadrupole moment and other relevant
collective variables are derived on the basis of the time dependent
Hartree-Fock-Bogoliubov equations. Analytical expressions for energy
centroids and transitions probabilities are found for the harmonic
oscillator model with the quadrupole-quadrupole residual interaction
and monopole pairing force. Deformation dependences of energies and
$B(M1)$ values are correctly reproduced. The inclusion of pair
correlations leads to a drastic improvement in the description of
qualitative and quantitative characteristics of the scissors mode.
\end{abstract}
      \vspace*{0.5cm}


\newpage

\section{Introduction}

An exhaustive analysis of the coupled dynamics of the scissors mode
and the isovector giant quadrupole resonance in a model of harmonic
oscillator with quadrupole--quadrupole residual interaction has been
performed in \cite{BaSc2}. The Wigner Function Moments (WFM) method
was applied to derive the
dynamical equations for angular momentum and quadrupole moment.
Analytical expressions for energies, $B(M1)$- and $B(E2)$-values, sum
rules and flow patterns of both modes were found for arbitrary values
of the deformation parameter. The subtle nature of the phenomenon and
its peculiarities were clarified.

Nevertheless, this description was not complete, because pairing was
not taken into account. It is well known \cite{Zaw}, that pairing is
very important for the correct quantitative description of the
scissors mode. Moreover, its role is crucial for an explanation of the
empirically observed deformation dependence of $E_{sc}$ and
$B(M1)_{sc}$.

The prediction of the scissors mode was inspired by the geometrical
picture of a counterrotating oscillation of the deformed proton
density against the deformed neutron density
\cite{Hilt92,Lo2000}. Thus, as it is seen from its physical nature,
the scissors mode can be observed only in deformed nuclei. Therefore,
quite naturally, the question of the deformation dependence of its
properties (for example, energy $E_{sc}$ and $B(M1)_{sc}$ value)
arises. However, during the first years after its discovery in
$^{156}$Gd \cite{Bohle} ``nearly all experimental data were limited to
nuclei of about the same deformation ($\delta\approx 0.20 - 0.25$),
and the important aspect of orbital $M1$ strength dependence on
$\delta$ has not yet been examined'', see ref. \cite{Ziegler}.

The first investigations of the $\delta$-dependence of $E_{sc}$ and
$B(M1)_{sc}$ were performed by W. Ziegler et al. \cite{Ziegler}, who
have studied the chain of isotopes $^{148,150,152,154}$Sm, and by
H. H. Pitz et al. \cite{Pitz} and J. Margraf et al.
\cite{Margraf}, who have studied the chain of isotopes
$^{142,146,148,150}$Nd. They found that the low-energy $B(M1)$ strength
exhibits approximately a quadratic dependence on the deformation
$\delta$.

Shortly afterwards it was discovered \cite{Ranga,Pietr1}, that in
even-even nuclei the total low-energy magnetic dipole strength is
closely related to the collective $E2$ strength of
the $2^+_1$ state and, thus, depends quadratically on the nuclear
deformation parameter.

Later J. Enders et al. \cite{Enders} made a theoretical
analysis of experimental data on the scissors mode in nuclei with
$140<A<200$. Investigating the sum rules $S_{+1}$ and $S_{-1}$
derived by E. Lipparini and S. Stringari \cite{Lipp,Strin}, they
found that the ratio $\bar\omega=S_{+1}/S_{-1}$ is proportional
to $E_{sc}$ with very good accuracy: $E_{sc}=0.44\bar\omega$. They
also observed that the moment of inertia $J_{gsb}$ of the ground
state rotational band and the moment of inertia for the irrotational
flow $J_{liq}=\delta^2J_{rig}$ (where $J_{rig}$ is the rigid body
moment if inertia) differ by nearly a constant factor
($K\approx10$) over the entire region. Using this fact and
identifying the giromagnetic ratio and the moment of inertia of the
scissors mode with those of the ground state rotational band, they
found with the help of the $S_{-1}$ sum rule, that $B(M1)_{sc}$ is
proportional to $\delta^2$.

So, all the rather numerous experimental data demonstrate
undoubtedly the $\delta^2$ dependence of $B(M1)_{sc}$ and the very
weak deformation dependence of $E_{sc}$. On the other hand, at the
beginning of the nuclear scissors studies all theoretical models,
starting from the first work by Suzuki and Rowe \cite{Suzuki},
predicted a linear $\delta$-dependence for both, $B(M1)_{sc}$ and
$E_{sc}$.

It turned out that the correct $\delta$-dependence is supplied
by the pairing correlations.
The effects of the pairing interaction in the description of the
scissors mode were evaluated for the first time by Bes and Broglia
\cite{Bes}. They assumed ``for simplicity that only the two subsets
of levels which are closest to the Fermi level ($n_{\perp}$ and
$n_{\perp}+1$) are affected by the pairing interactions''. In this case
$B(M1)$ should be multiplied by the factor
$(u_{n_{\perp}}v_{n_{\perp}+1}-u_{n_{\perp}+1}v_{n_{\perp}})^2\approx$
$(u_{n_{\perp}}^2-v_{n_{\perp}}^2)^2=(e_{n_{\perp}}/E_{n_{\perp}})^2$
with $E_i=\sqrt{e_i^2+\Delta^2}$ and $e_i=\epsilon_i-\mu$, where
$\epsilon_i$ is a single particle energy, $\Delta$ is the gap and $\mu$
is the chemical
potential. The value of $(e_{n_{\perp}}/E_{n_{\perp}})$ was found by
``making use of the fact that the moment of inertia is approximately
1/2 of the rigid body value obtained in the absence of pairing''. Thus,
in accordance with the Inglis formula one has
\begin{equation}
1/2=J/J_{rig}=[(u_{n_{\perp}}^2-v_{n_{\perp}}^2)^2/2E_{n_{\perp}}]/
[1/2e_{n_{\perp}}]=(e_{n_{\perp}}/E_{n_{\perp}})^3
\label{e/E}
\end{equation}
and $e_{n_{\perp}}/E_{n_{\perp}}$=0.79. As a result $B(M1)$ is reduced
by the factor $(e_{n_{\perp}}/E_{n_{\perp}})^2$ = 0.62, i.e. the
influence of pairing is quite remarkable.

It was, however, noted by Hamamoto and Magnusson \cite{Magnus} that
this result holds only for well deformed nuclei, where the
equality (\ref{e/E}) is valid. In general it is necessary to
take into account the $\delta$-dependence of the
$e_{n_{\perp}}/E_{n_{\perp}}$ - factor. This was done for the first
time in ref. \cite{Magnus}. The authors applied
``the method of averaging the position of the chemical potential
between the occupied subshell $(N,n_{\perp})$ and the empty shell
$(N,n_{\perp}+1)$'' to find that the $\delta$-dependence of $B(M1)$ is
determined by the function
\begin{equation}
Y=(\delta A^{4/3})\left( \sqrt{1+4x^2}
+\frac{1}{2x}\ln|\sqrt{1+4x^2}+2x|\right)\frac{1}{x}
\left(1-\frac{1}{x\sqrt{1+x^2}}\ln|\sqrt{1+x^2}+x|\right)
\label{Hamam}
\end{equation}
with $x=(\hbar\omega_0\delta)/(2\Delta)$. In the small deformation limit
this function is proportional to $\delta^2$, while for large $\delta$ it
deviates remarkably from such a simple dependence. The authors
performed
also a more realistic QRPA calculation for the Woods-Saxon potential
with QQ and $\sigma\sigma$ residual interactions, which confirmed
their simplified analytical estimate.

In \cite{Pietr}, N. Pietralla et al. established
the $\delta$-dependence of the
$e_{n_{\perp}}/E_{n_{\perp}}$ - factor phenomenologically. They were
first to perform the theoretical analysis of the experimental data
of the scissors mode in nuclei in the mass region $130<A<200$.
Following the idea of Bes and Broglia they
parametrized the $e_{n_{\perp}}/E_{n_{\perp}}$ - factor as
$$E_{n_{\perp}}/e_{n_{\perp}}=\sqrt{1+(b\delta)^2}/(a\delta).$$
The free parameters $a$ and $b$ were fixed by a fit to the
experimental moments of inertia with the help of a
formula equivalent to (\ref{e/E})
\begin{equation}
(e_{n_{\perp}}/E_{n_{\perp}})^3=J_{exp}/J_{rig},
\label{e/E3}
\end{equation}
where $J_{exp}=3\hbar^2/E(2_1^+)$ is the effective moment of inertia
of the ground state band. In this way it was found that
``the centers of
gravity of the observed $M1$ strength distributions are always close to
3 MeV'', i.e. ``the data exhibit a weak dependence of the scissors mode
on the deformation parameter''. They also derived a semiempirical
formula for the total $M1$ strength of the scissors mode
\begin{equation}
B(M1;0_1^+\to 1_{sc}^+)=2.6c_g^2\frac{\delta^3}{1+(3\delta)^2}
\frac{Z^2}{A^{2/3}}\mu_N^2
\label{BMPietr}
\end{equation}
($c_g=0.8$ is the scaling factor of the giromagnetic ratio), which
describes very well the experimental data and gives a deformation
dependence ``practically indistinguishable from the $\delta^2$
dependence''.

A direct way to demonstrate the $\delta$-dependence of the
$e_{n_{\perp}}/E_{n_{\perp}}$ - factor
was suggested in \cite{Garrido}. E. Garrido et al. have
shown that it is possible to extract analytically the $\delta^2$ factor
from the occupation coefficient
$\Phi_{\alpha\beta}=(u_{\alpha}v_{\beta}-u_{\beta}v_{\alpha})^2.$
Using the definitions of the $u_{\alpha}, v_{\alpha}$
coefficients, it is easy to write $\Phi_{\alpha\beta}$ as
\begin{equation}
\Phi_{\alpha\beta}=\frac{(e_{\alpha}-e_{\beta})^2}
{4E_{\alpha}E_{\beta}}(1-P_{\alpha\beta})
\label{Phiab}
\end{equation}
with
\begin{equation}
P_{\alpha\beta}=\frac{(E_{\alpha}-E_{\beta})^2}
{(e_{\alpha}-e_{\beta})^2}
=\frac{1}{2E^2}[z-(z^2-4e^2_{\alpha\beta}E^2)^{1/2}]
=\frac{e^2_{\alpha\beta}}{z}[1+\frac{e^2_{\alpha\beta}E^2}{z^2}+...]
\label{Pab}
\end{equation}
and $E=\frac{1}{2}(e_{\alpha}-e_{\beta}),
e_{\alpha\beta}=\frac{1}{2}(e_{\alpha}+e_{\beta})$,
$z=e_{\alpha\beta}^2+E^2+\Delta^2
=\frac{1}{2}(e_{\alpha}^2+e_{\beta}^2)+\Delta^2
=\frac{1}{2}(E_{\alpha}^2+E_{\beta}^2)$.
For the scissors mode $e_{\alpha}-e_{\beta}\simeq\hbar\omega\delta$.
The function $(1-P_{\alpha\beta})/(E_{\alpha}E_{\beta})$ has a
regular dependence on $\delta$ (no poles), so the coefficient
$\Phi_{\alpha\beta}$ and, respectively, $B(M1)_{sc}$ are obviously
proportional to $\delta^2$. For the IVGQR
$e_{\alpha}-e_{\beta}\simeq2\hbar\omega$, but its $B(M1)$ is
proportional to $\delta^2$ even without pairing due to other reasons
(see section 3).

In this paper
we generalize the WFM method to take into account pair
correlations. This allows us to obtain the correct $\delta$-dependence
for $E_{sc}$ and $B(M1)_{sc}$ in a slightly different way than in the
papers, cited above.

The paper is organized as follows. In section 2 the moments of
Time Dependent Hartree-Fock-Bogoliubov (TDHFB)
dynamical equations for normal and abnormal densities are calculated
and adequate approximations are introduced to obtain the
final set of six dynamical equations for the collective variables. In
section 3 these equations are decoupled in the isoscalar and isovector
sets and the isovector excitation energies and transitions
probabilities are calculated in the framework of the HO+QQ model. The
results of calculations are discussed in section 4.
Concluding remarks are contained in section 5. Some mathematical
details are given in Appendices.

\section{Phase space moments of TDHFB equations}

The time dependent HFB equations in matrix formulation are
\cite{Solov,Ring}
\begin{equation}
i\hbar\dot\R=[\H,\R]
\label{tHFB}
\end{equation}
with
$$
\R={\hat\rho\qquad-\hat\kappa\choose-\hat\kappa^{\dagger}\;\;1-\hat\rho^*},
\quad\H={\hat h\quad\;\;\hat\Delta\choose\hat\Delta^{\dagger}\quad-\hat h^*}
$$
The normal density matrix $\hat \rho$ and Hamiltonian $\hat h$ are hermitian;
the abnormal density $\hat \kappa$ and the pairing gap $\hat \Delta$ are skew
symmetric
$$\hat \kappa^{\dagger}=-\hat \kappa^*,\quad
\hat \Delta^{\dagger}=-\hat \Delta^*.$$

The detailed form of the HFB equations is
\begin{eqnarray}
i\hbar\dot{\hat\rho} =\hat h\hat\rho -\hat\rho\hat h
-\hat\Delta \hat\kappa ^{\dagger}+\hat\kappa \hat\Delta^\dagger,&&
-i\hbar\dot{\hat\kappa} =-\hat h\hat\kappa -\hat\kappa \hat h^*+\hat\Delta
-\hat\Delta \hat\rho ^*-\hat\rho \hat\Delta ,
\nonumber\\
-i\hbar\dot{\hat\rho}^*=\hat h^*\hat\rho ^*-\hat\rho ^*\hat h^*
-\hat\Delta^\dagger\hat\kappa +\hat\kappa^\dagger\hat\Delta ,&&
-i\hbar\dot{\hat\kappa}^\dagger=\hat h^*\hat\kappa^\dagger
+\hat\kappa^\dagger\hat h-\hat\Delta^\dagger
+\hat\Delta^\dagger\hat\rho +\hat\rho^*\hat\Delta^\dagger .
\label{HFB}
\end{eqnarray}
We will work with the Wigner transformation \cite{Ring} of these
equations. The relevant
mathematical details can be found in Appendix A.
To make the formulae more transparent, in the following we will not
specify the spin and isospin indices. The isospin indices will be
re-introduced at the end.
As a rule, we also will not write out the coordinate dependence
$(\br,\bp)$ of all functions.
The Wigner transform of (\ref{HFB}) can be written as
\begin{eqnarray}
i\hbar\dot f &=&i\hbar\{h,f\}
-\Delta\kappa^{*}+\kappa\Delta^{*}
-\frac{i\hbar}{2}\{ \Delta,\kappa^{*}\}
+\frac{i\hbar}{2}\{ \kappa,\Delta^{*}\}
\nonumber\\
&&-
\frac{\hbar^2}{8}
[\{\{ \kappa,\Delta^{*}\}\}-\{\{ \Delta,\kappa^{*}\}\}]+...,
\nonumber\\
-i\hbar\dot{\bar f}&=&i\hbar\{\bar h,\bar f\}
-\Delta^{*}\kappa+\kappa^{*}\Delta
-\frac{i\hbar}{2}\{ \Delta^{*},\kappa\}
+\frac{i\hbar}{2}\{ \kappa^{*},\Delta\}
\nonumber\\
&&
+\frac{\hbar^2}{8}
[\{\{ \kappa,\Delta^{*}\}\}-\{\{ \Delta,\kappa^{*}\}\}]+...,
\nonumber\\
-i\hbar\dot\kappa&=&
-h\kappa-\kappa\bar h
-\frac{i\hbar}{2}\{h,\kappa\}
-\frac{i\hbar}{2}\{\kappa,\bar h\}
\nonumber\\
&&
+\Delta
-\Delta\bar f-f\Delta
-\frac{i\hbar}{2}\{f,\Delta\}
-\frac{i\hbar}{2}\{\Delta,\bar f\}
\nonumber\\
&&+\frac{\hbar^2}{8}
[\{\{h,\kappa\}\}+\{\{\kappa,\bar h\}\}
+\{\{ \Delta,\bar f\}\}+\{\{f,\Delta\}\}]+...,
\nonumber\\
-i\hbar\dot\kappa^{*}&=&
\kappa^{*}h+\bar h\kappa^{*}
+\frac{i\hbar}{2}\{\kappa^{*},h\}
+\frac{i\hbar}{2}\{\bar h ,\kappa^{*}\}
\nonumber\\
&&
-\Delta^{*}+\bar f\Delta^{*}+\Delta^{*} f
+\frac{i\hbar}{2}\{\bar f,\Delta^{*}\}
+\frac{i\hbar}{2}\{\Delta^{*},f\}
\nonumber\\
&&
-\frac{\hbar^2}{8}
[\{\{\kappa^{*},h\}\}+\{\{\bar h ,\kappa^{*}\}\}
+\{\{\bar f,\Delta^{*}\}\}+\{\{\Delta^{*},f\}\}]+... ,
\label{WHFB}
\end{eqnarray}
where the functions $h$, $f$, $\Delta$, and $\kappa$ are the Wigner
transforms of $\hat h$, $\hat\rho$, $\hat\Delta$, and $\hat\kappa$,
respectively, $\bar f(\br,\bp)=f(\br,-\bp)$, $\{f,g\}$ is the Poisson
bracket of the functions $f(\br,\bp)$ and $g(\br,\bp)$ (see Appendix A);
the dots stand for terms proportional to higher powers of $\hbar$.

To investigate collective modes described by these equations we apply
the method of Wigner function moments. The idea of the method is based
on the virial theorems by Chandrasekhar and Lebovitz \cite{Chand}; its
detailed formulation can be found in \cite{Bal,BaSc}. To study the
quadrupole collective motion in axially symmetric nuclei it is
necessary to calculate moments of Eqs.~(\ref{WHFB}) with the
weight functions
\begin{equation}
xz,\quad p_xp_z,\quad zp_x+xp_z,\quad \mbox{and}\quad zp_x-xp_z.
\end{equation}
This procedure means that we refrain from seeking the whole density
matrix and restrict
ourselves to the knowledge of only several moments. Nevertheless this
information turns out to be sufficient for a satisfactory description
of various collective modes with quantum numbers $K^{\pi}=1^+$, as it
was shown in our
previous publications \cite{BaSc2,Bal,BaSc}. In the case without
pairing, this restricted information can be extracted from the TDHF
equations and becomes exact only for the harmonic oscillator with
multipole-multipole residual interactions. For more realistic models
it becomes approximate even without pairing. The TDHFB equations
(\ref{WHFB}) are considerably more complicated than the TDHF ones, so
additional approximations are necessary even for the simple model
considered here. This is the subject of this section.

Let us at first write out several useful relations:
$$
\int\! d^3p\int\! d^3r
 A\{f,g\}=-
\int\! d^3p\int\! d^3r
f\{A,g\}=-
\int\! d^3p\int\! d^3r
g\{f,A\},$$
$$
\int\! d^3p\int\! d^3r
A\{\{f,g\}\}=
\int\! d^3p\int\! d^3r
f\{\{A,g\}\}=
\int\! d^3p\int\! d^3r
g\{\{f,A\}\},
$$
where $A$ is any one of the above mentioned weight functions, $f$ and
$g$ are arbitrary functions and
$\{\{f,g\}\}$ is defined in Appendix B.
Integration of Eqs.~(\ref{WHFB}) (including the terms of
higher orders in $\hbar$) over the phase space with the weight $A$
yields the following set of equations:
\begin{eqnarray}
      i\hbar\frac{d}{dt}\int\! d(\bp,\br) A f=
\int\! d(\bp,\br)
\left[i\hbar \{A,h\}f+A(\Delta^{*}\kappa-\kappa^{*}\Delta)
-\frac{i\hbar}{2}(\{A,\Delta^{*}\}\kappa+\{A,\Delta\}\kappa^{*})
\right.
\nonumber\\
\left.
-\frac{\hbar^2}{8}
(\{\{A,\Delta^{*}\}\}\kappa-\{\{A,\Delta\}\}\kappa^{*})
\right],
\nonumber\\
     i\hbar\frac{d}{dt}\int\! d(\bp,\br) A\bar f=
\int\! d(\bp,\br)
\left[i\hbar\{\bar h,A\}\bar f+A(\Delta^{*}\kappa-\kappa^{*}\Delta)
+\frac{i\hbar}{2}(\{A,\Delta^{*}\}\kappa+\{A,\Delta\}\kappa^{*})
\right.
\nonumber\\
\left.
-\frac{\hbar^2}{8}
(\{\{A,\Delta^{*}\}\}\kappa-\{\{A,\Delta\}\}\kappa^{*})
\right],
\nonumber\\
     i\hbar\frac{d}{dt}\int\! d(\bp,\br) A\kappa=
\int\! d(\bp,\br)\left[A(h+\bar h)\kappa
+\frac{i\hbar}{2}\{A,(h-\bar h)\}\kappa
-A\Delta(1-\bar f-f)
\right.
\hspace{16mm}
\nonumber\\
\left.
+\frac{i\hbar}{2}\{A,\Delta\}(\bar f-f)
-\frac{\hbar^2}{8}
[\{\{A,(h+\bar h)\}\}\kappa
+\{\{A,\Delta\}\}(\bar f+f)]
\right],
\nonumber\\
     i\hbar\frac{d}{dt}\int\! d(\bp,\br) A\kappa^{*}=
\int\! d(\bp,\br)\left[-A(h+\bar h)\kappa^{*}
+\frac{i\hbar}{2}\{A,(h-\bar h)\}\kappa^{*}
+A\Delta^{*}(1-\bar f-f)
\right.
\nonumber\\
\left.
+\frac{i\hbar}{2}\{A,\Delta^{*}\}(\bar f-f)
+\frac{\hbar^2}{8}
[\{\{A,(h+\bar h)\}\}\kappa^{*}
+\{\{A,\Delta^{*}\}\}(\bar f+f)]
\right],
\label{HFBA}
\end{eqnarray}
where $\int\! d(\bp,\br)\equiv
2(2\pi\hbar)^{-3}\int\!d^3p\,\int\!d^3r$.  It is necessary to note an
essential point: there are no terms with higher powers of $\hbar$ in
these equations. The infinite number of terms proportional to
$\hbar^n$ with $n>2$ have disappeared after integration, as is
demonstrated in Appendix B. This fact does not mean, that higher
powers of $\hbar$ are not necessary for the exact solution of the
problem. As it will be shown below, the set of equations (\ref{HFBA})
contains terms, which couple with dynamical equations of
higher order moments, which include, naturally, the higher powers of
$\hbar$.

It is convenient to rewrite the above equations in terms of
$h_{\pm}=h\pm \bar h$, $f_{\pm}=f\pm\bar f$,
$\Delta_{\pm}=\Delta\pm\Delta^*$, $\kappa_{\pm}=\kappa\pm\kappa^*$:
\begin{eqnarray}
      i\hbar\frac{d}{dt}\int\! d(\bp,\br) A f_+&=&
\int\! d(\bp,\br)\left[\frac{i\hbar}{2}
(\{A,h_+\}f_-+\{A,h_-\}f_+)
+A(\Delta_+\kappa_--\kappa_+\Delta_-)
\right.
\nonumber\\
&&\left.
-\frac{\hbar^2}{8}
(\{\{A,\Delta_+\}\}\kappa_--\{\{A,\Delta_-\}\}\kappa_+)
\right],
\nonumber\\
      i\hbar\frac{d}{dt}\int\! d(\bp,\br) A f_-&=&
\int\! d(\bp,\br)\left[\frac{i\hbar}{2}
(\{A,h_+\}f_++\{A,h_-\}f_-)
\right.
\nonumber\\
&&\left.
-\frac{i\hbar}{2}(\{A,\Delta_+\}\kappa_+
-\{A,\Delta_-\}\kappa_-)
\right],
\nonumber\\
i\hbar\frac{d}{dt}\int\! d(\bp,\br) A\kappa_+&=&
\int\! d(\bp,\br)\left[Ah_+\kappa_-
+\frac{i\hbar}{2}\{A,h_-\}\kappa_+
-A\Delta_-(1-f_+)
\right.
\nonumber\\
&&\left.
-\frac{i\hbar}{2}\{A,\Delta_+\}f_-
-\frac{\hbar^2}{8}
(\{\{A,h_+\}\}\kappa_-
+\{\{A,\Delta_-\}\}f_+)
\right],
\nonumber\\
i\hbar\frac{d}{dt}\int\! d(\bp,\br) A\kappa_-&=&
\int\! d(\bp,\br)\left[Ah_+\kappa_+
+\frac{i\hbar}{2}\{A,h_-\}\kappa_-
-A\Delta_+(1-f_+)
\right.
\nonumber\\
&&\left.-\frac{i\hbar}{2}\{A,\Delta_-\}f_-
-\frac{\hbar^2}{8}(\{\{A,h_+\}\}\kappa_+
+\{\{A,\Delta_+\}\}f_+)
\right].
\label{HFBA+}
\end{eqnarray}

These equations are strongly nonlinear, because $\Delta$ is a function of
$\kappa$ (see, e.g., ref. \cite{Ring}):
\begin{equation}
\Delta(\br,\bp)=\int\! \frac{d^3p'}{(2\pi\hbar)^3}
v(|\bp-\bp'|)\kappa(\br,\bp').
\label{DK}
\end{equation}
Having in mind small amplitude oscillations we will linearize:
$f_{\pm}=f_{\pm}^{0}+\delta f_{\pm}$,
$\kappa_{\pm}=\kappa^{0}_{\pm}+\delta\kappa_{\pm}$,
$\Delta_{\pm}=\Delta^{0}_{\pm}+\delta\Delta_{\pm}$.
The Hamiltonian should be divided into the ground state Hamiltonian
$h^{0}$ and the residual interaction $h^{1}$ (and, if necessary, the
external field). We consider
$h^{0}$ without $\bp$-odd terms, hence $h^{0}_-=0$ and as a
consequence $f_-^{0}=0$. It is natural to take $\Delta^{0}$ real, i.e.
$\Delta^{0}_-=0$, $\kappa^{0}_-=0$.
Linearizing (\ref{HFBA+}) and taking into account the last remarks we
arrive at
\begin{eqnarray}
&&      i\hbar\frac{d}{dt}\int\! d(\bp,\br) A \delta f_+=
\int\! d(\bp,\br)\left[
\frac{i\hbar}{2}(
\{A,h_{+}^{0}\}\delta f_-
+\{A,h_{-}^{1}\}f_+^{0})
+A(\Delta_+^{0}\delta \kappa_--\kappa_+^{0}\delta \Delta_-)
\right.\hspace*{5mm}
\nonumber\\
&&\hspace*{40mm}
\left.
-\frac{\hbar^2}{8}
(\{\{A,\Delta_+^{0}\}\}\delta \kappa_-
-\{\{A,\delta \Delta_-\}\}\kappa_+^{0})
\right],
\nonumber\\
&&      i\hbar\frac{d}{dt}\int\! d(\bp,\br) A \delta f_-=
\int\! d(\bp,\br)\left[
\frac{i\hbar}{2}(
\{A,h_{+}^{0}\}\delta f_+
+\{A,h_{+}^{1}\}f_+^{0})
\right.
\nonumber\\
&&\hspace*{40mm}
\left.
-\frac{i\hbar}{2}(\{A,\Delta_+^{0}\}\delta \kappa_+
+\{A,\delta \Delta_+\}\kappa_+^{0})
\right],
\nonumber\\
&&i\hbar\frac{d}{dt}\int\! d(\bp,\br) A\delta \kappa_+=
\int\! d(\bp,\br)\left[Ah_+^{0}\delta \kappa_-
+\frac{i\hbar}{2}\{A,h_-^{1}\}\kappa_+^{0}
-A\delta \Delta_-(1-f_+^{0})
\right.
\nonumber\\
&&\hspace*{40mm}
\left.
-\frac{i\hbar}{2}\{A,\Delta_+^{0}\}\delta f_-
-\frac{\hbar^2}{8}
(\{\{A,h_+^{0}\}\}\delta \kappa_-
+\{\{A,\delta \Delta_-\}\}f_+^{0})
\right],
\nonumber\\
&&i\hbar\frac{d}{dt}\int\! d(\bp,\br) A\delta \kappa_-=
\int\! d(\bp,\br)\left[Ah_+^{0}\delta \kappa_+
+Ah_+^{1}\kappa_+^{0}
-A\delta \Delta_+(1-f_+^{0})
+A\Delta_+^{0}\delta f_+
\right.
\nonumber\\
&&\hspace*{20mm}
\left.
-\frac{\hbar^2}{8}(\{\{A,h_+^{0}\}\}\delta \kappa_+
+\{\{A,h_+^{1}\}\}\kappa_+^{0}
+\{\{A,\Delta_+^{0}\}\}\delta f_+
+\{\{A,\delta \Delta_+\}\}f_+^{0})
\right].
\label{HFBAlin}
\end{eqnarray}
Until this point, our formulation is completely general.

Now let us consider the popular case of pure monopole pairing. This
means that the variation of the gap,
\begin{equation}
\delta\Delta(\br,\bp)=\int\! \frac{d^3p'}{(2\pi\hbar)^3}
v(|\bp-\bp'|)\delta\kappa(\br,\bp'),
\label{varDK}
\end{equation}
will be projected on its monopole part. In the case of quadrupole
vibrations, which we will study here, the variations $\delta f_{\pm}$
and $\delta\kappa_{\pm}$ will have quadrupole multipolarities. As a
consequence, when projecting formula (\ref{varDK}) on the monopole
part, the integral over angles will be equal to zero and we get
$\delta \Delta=0$. We also note that neglecting $\delta\Delta$
corresponds to the usual Inglis approximation. Then
Eqs.~(\ref{HFBAlin}) are reduced to
\begin{eqnarray}
      i\hbar\frac{d}{dt}\int\! d(\bp,\br) A \delta f_+&=&
\int\! d(\bp,\br)\left[
\frac{i\hbar}{2}(
\{A,h_{+}^{0}\}\delta f_-
+\{A,h_{-}^{1}\}f_+^{0})
\right.
\nonumber\\
&&\left.
+A\Delta_+^{0}\delta \kappa_-
-\frac{\hbar^2}{8}
\{\{A,\Delta_+^{0}\}\}\delta \kappa_-
\right],
\nonumber\\
      i\hbar\frac{d}{dt}\int\! d(\bp,\br) A \delta f_-&=&
\int\! d(\bp,\br)\left[
\frac{i\hbar}{2}(
\{A,h_{+}^{0}\}\delta f_+
+\{A,h_{+}^{1}\}f_+^{0})
-\frac{i\hbar}{2}\{A,\Delta_+^{0}\}\delta \kappa_+
\right],
\nonumber\\
i\hbar\frac{d}{dt}\int\! d(\bp,\br) A\delta \kappa_+&=&
\int\! d(\bp,\br)\left[Ah_+^{0}\delta \kappa_-
+\frac{i\hbar}{2}\{A,h_-^{1}\}\kappa_+^{0}
\right.
\nonumber\\
&&\left.
-\frac{i\hbar}{2}\{A,\Delta_+^{0}\}\delta f_-
-\frac{\hbar^2}{8}
\{\{A,h_+^{0}\}\}\delta \kappa_-
\right],
\nonumber\\
i\hbar\frac{d}{dt}\int\! d(\bp,\br) A\delta \kappa_-&=&
\int\! d(\bp,\br)\left[Ah_+^{0}\delta \kappa_+
+Ah_+^{1}\kappa_+^{0}
+A\Delta_+^{0}\delta f_+
\right.
\nonumber\\
&&\left.
-\frac{\hbar^2}{8}
(\{\{A,h_+^{0}\}\}\delta \kappa_+
+\{\{A,h_+^{1}\}\}\kappa_+^{0}
+\{\{A,\Delta_+^{0}\}\}\delta f_+)
\right].
\label{HFBAd0}
\end{eqnarray}

To proceed further we are forced to do two approximations to get rid
of higher rank moments and obtain a closed set of dynamical equations
for second rank moments. First, the integrals $\int\! d(\bp,\br)A
h_+^{0}\delta \kappa_{\pm}$ contain fourth rank moments. The analysis
of the integrand shows that we can neglect these integrals without a
strong loss of accuracy. Indeed, the functions $\kappa_{\pm}$ (and
their variations) are sharply peaked at the Fermi surface, where the
Hamiltonian $h_+^{0}$ by definition is equal to zero. Therefore the
product $h_+^{0}\delta \kappa_{\pm}$ should be small. Second, the
($\br,\bp$)-dependence of $\Delta^{0}_+$ can generate, in principle,
an infinite number of moments of various ranks. To simplify the
problem we will consider here the commonly employed approximation of an
($\br,\bp$)-independent gap $\Delta^{0}_+\equiv 2\Delta=\mathit{const}$,
an approximation often used in nuclear physics and consistent with the
monopole-monopole pairing force model. So, adding the isospin index
$\tau$ = (p,n), we finally have
\begin{eqnarray}
      i\hbar\frac{d}{dt}\int\! d(\bp,\br) A \delta f_+^{\tau}&=&
\int\! d(\bp,\br)\left[
\frac{i\hbar}{2}(
\{A,h_{+}^{\tau0}\}\delta f_-^{\tau}
+\{A,h_{-}^{\tau1}\}f_+^{\tau0})
+2A\Delta^{\tau}\delta \kappa_-^{\tau}
\right],
\nonumber\\
      i\hbar\frac{d}{dt}\int\! d(\bp,\br) A \delta f_-^{\tau}&=&
\int\! d(\bp,\br)\left[
\frac{i\hbar}{2}(
\{A,h_{+}^{\tau0}\}\delta f_+^{\tau}
+\{A,h_{+}^{\tau1}\}f_+^{\tau0})
\right],
\nonumber\\
i\hbar\frac{d}{dt}\int\! d(\bp,\br) A\delta \kappa_+^{\tau}&=&
\int\! d(\bp,\br)\left[
\frac{i\hbar}{2}\{A,h_-^{\tau1}\}\kappa_+^{\tau0}
-\frac{\hbar^2}{8}
\{\{A,h_+^{\tau0}\}\}\delta \kappa_-^{\tau}
\right],
\nonumber\\
i\hbar\frac{d}{dt}\int\! d(\bp,\br) A\delta \kappa_-^{\tau}&=&
\int\! d(\bp,\br)\left[Ah_+^{\tau1}\kappa_+^{\tau0}
+2A\Delta^{\tau}\delta f_+^{\tau}
\right.
\nonumber\\
&&\left.
-\frac{\hbar^2}{8}
(\{\{A,h_+^{\tau0}\}\}\delta \kappa_+^{\tau}
+\{\{A,h_+^{\tau1}\}\}\kappa_+^{\tau0})
\right].
\label{HFBAdcon}
\end{eqnarray}

We consider an axially symmetric model with
$\di h_0^{\tau}=\frac{p^2}{2m}+\frac{1}{2}m
[\omega_x^{\tau2}(x^2+y^2)+\omega_z^{\tau2}z^2]-\mu^{\tau}$ and
$h_1^{\tau}=Z^{\tau}(t)xz$ with
$Z^{\rm n}=\chi_{nn}Q^{\rm n}+\chi_{np}Q^{\rm p}$ and
$Z^{\rm p}=\chi_{pp}Q^{\rm p}+\chi_{np}Q^{\rm n}$, $\mu^{\tau}$ being
the chemical potential of either protons ($\tau$=p) or neutrons
($\tau$=n).
$\chi_{\tau\tau'}$ is the strength constant
of the quadrupole-quadrupole residual interaction, $Q^{\tau}$ is a
component of the quadrupole moment
$Q^{\tau}(t)=\int\! d(\bp,\br) xz \delta f^{\tau}(\br,\bp,t)$.
It is supposed that $\chi_{nn}=\chi_{pp}$. Obviously, in this
model $h_-^{\tau}=0$ and $h_+^{\tau}=2h^{\tau}$.
Calculating the required Poisson brackets
$$\{xz,h_0^{\tau}\}=\frac{1}{m}\hat L,\quad
\{p_xp_z,h_0^{\tau}\}=-m(\omega_z^{\tau2}zp_x+\omega_x^{\tau2}xp_z),
\quad \{zp_x,h_0^{\tau}\}=\frac{1}{m}p_xp_z-m\omega_x^{\tau2}xz,
$$
$$
\{xp_z,h_0^{\tau}\}=\frac{1}{m}p_xp_z-m\omega_z^{\tau2}xz,
\quad\{xz,h_1^{\tau}\}=0,\quad
\{p_xp_z,h_1^{\tau}\}=-Z^{\tau}(t)(zp_z+xp_x),
$$
$$
\{zp_x,h_1^{\tau}\}=-Z^{\tau}(t)z^2, \quad
\{xp_z,h_1^{\tau}\}=-Z^{\tau}(t)x^2, \quad \{\{A,h_0^{\tau}\}\}=0,
$$
$$
\{\{xz,h_1^{\tau}\}\}=0,\quad
\{\{p_xp_z,h_1^{\tau}\}\}=2Z^{\tau}(t),\quad
\{\{zp_x,h_1^{\tau}\}\}=0, \quad \{\{xp_z,h_1^{\tau}\}\}=0.
$$
we find, that the third equation of (\ref{HFBAdcon}) becomes trivial,
giving four integrals of motion\\
 $\int\! d(\bp,\br) A\delta \kappa_+^{\tau}$.
Introducing the notation
$$
Q^{\tau}(t)=\int\! d(\bp,\br) xz \delta
f^{\tau}(\br,\bp,t),\quad
\tilde Q^{\tau}(t)=\int\! d(\bp,\br) xz \delta\kappa^{\tau}_-(\br,\bp,t),$$
$$
P^{\tau}(t)=
\int\! d(\bp,\br) p_xp_z \delta
f^{\tau}(\br,\bp,t),\quad
\tilde P^{\tau}(t)=\int\! d(\bp,\br) p_xp_z \delta
\kappa^{\tau}_-(\br,\bp,t),$$
$$
L^{\tau}(t)=
\int\! d(\bp,\br)(zp_x+xp_z)
\delta f^{\tau}(\br,\bp,t),\quad
\tilde L^{\tau}(t)=\int\! d(\bp,\br)(zp_x+xp_z) \delta
\kappa^{\tau}_-(\br,\bp,t),$$
$$
I_{y}^{\tau}(t)=
\int\! d(\bp,\br)(zp_x-xp_z) \delta
f^{\tau}(\br,\bp,t),\quad
\tilde I_y^{\tau}(t)=\int\! d(\bp,\br)(zp_x-xp_z)
\delta\kappa^{\tau}_-(\br,\bp,t)$$
we find the following set of dynamical equations
\begin{eqnarray}
i\hbar \dot Q^{\tau}&=&i\hbar \frac{1}{m}L^{\tau}
+\Delta^{\tau}\tilde Q^{\tau},
\nonumber\\
i\hbar \dot{\tilde Q}^{\tau}&=&4\Delta^{\tau} Q^{\tau}
+2k_4^{\tau}Z^{\tau},
\nonumber\\
i\hbar \dot P^{\tau}&=&-i\hbar \frac{m}{2}[(\omega_x^{\tau2}
+\omega_z^{\tau2})L^{\tau}
-(\omega_x^{\tau2}-\omega_z^{\tau2})I_y^{\tau}]
+\Delta^{\tau}\tilde P^{\tau},
\nonumber\\
i\hbar \dot{\tilde P}^{\tau}&=&4\Delta^{\tau} P^{\tau}
-\frac{\hbar^2}{2} k_0^{\tau} Z^{\tau},
\nonumber\\
i\hbar \dot L^{\tau}&=&i\hbar [\frac{2}{m}P^{\tau}
-m(\omega_x^{\tau2}+\omega_z^{\tau2})Q^{\tau}
-(\langle z^2\rangle^{\tau}+\langle x^2\rangle^{\tau})
Z^{\tau}],
\nonumber\\
i\hbar \dot I_y^{\tau}&=&
i\hbar m(\omega_z^{\tau2}-\omega_x^{\tau2})Q^{\tau}
-i\hbar
(\langle z^2\rangle^{\tau}-\langle x^2\rangle^{\tau})Z^{\tau}.
\label{HFB6}
\end{eqnarray}
where
\begin{eqnarray}
\langle x^2\rangle^{\tau}=
\int\! d(\bp,\br)x^2f^{\tau0}(\br,\bp),&&\quad
\langle z^2\rangle^{\tau}=
\int\! d(\bp,\br)z^2f^{\tau0}(\br,\bp),
\nonumber\\
k_0^{\tau}=\int\! d(\bp,\br) \kappa_+^{\tau0}(\br,\bp),&&\quad
k_4^{\tau}=\int\! d(\bp,\br)x^2z^2 \kappa_+^{\tau0}(\br,\bp).
\label{k0k4}
\end{eqnarray}
Eqs. (\ref{HFB6}) will be simplified as far as possible to obtain
results in analytical form.

\section{Simplified model}

The scissors mode is an isovector one, so it is natural to rewrite
Eqs.~(\ref{HFB6}) in isoscalar and isovector terms. For the
scissors mode, which we are interested in, the isovector set of
equations can be decoupled from the isoscalar one with the help of the
following approximations:

$$
\Delta^{\rm p}\simeq \Delta^{\rm n}=\Delta,\quad
k_4^{\rm p}\simeq k_4^{\rm n}=k_4/2,\quad
k_0^{\rm p}\simeq k_0^{\rm n}=k_0/2,
$$
$$
\langle x^2\rangle^{\rm p}\simeq\langle x^2\rangle^{\rm n}=
\langle x^2\rangle/2,\quad
\langle z^2\rangle^{\rm p}\simeq \langle z^2\rangle^{\rm n}=
\langle z^2\rangle/2,\quad
\delta^{\rm p}\simeq\delta^{\rm n}=\delta,
$$
where $\delta$ is the nucleus deformation. Introducing isovector
variables $\underline Q=Q^{\rm n}-Q^{\rm p}$, $\underline{\tilde
Q}=\tilde Q^{\rm n}-\tilde Q^{\rm p}$ and so on, we can write the
isovector set of equations as
\begin{eqnarray}
i\hbar \underline{\dot Q}&=&i\hbar \frac{1}{m}\underline{L}
+\Delta\underline{\tilde Q},
\nonumber\\
i\hbar \underline{\dot{\tilde Q}}&=&4\Delta \underline{Q}
+2k_4\chi_1 \underline{Q},
\nonumber\\
i\hbar \underline{\dot P}&=&-i\hbar m\bar{\omega}^2[(1+\frac{\delta}{3})
\underline{L}-\delta \underline{I_y}]+\Delta\underline{\tilde P},
\nonumber\\
i\hbar \underline{\dot{\tilde P}}&=&4\Delta \underline{P}
-\frac{\hbar^2}{2} k_0 \chi_1 \underline{Q},
\nonumber\\
i\hbar \underline{\dot L}&=&i\hbar [\frac{2}{m}\underline{P}
-(2m\bar\omega^2+\frac{2}{3}Q_{00}\chi_1)
(1+\frac{\delta}{3})\underline{Q}],
\nonumber\\
i\hbar \underline{\dot I_y}&=&
-i\hbar\delta(2m\bar\omega^2+\frac{2}{3}Q_{00}\chi_1)
\underline{Q},
\label{Ivect}
\end{eqnarray}
where $\chi_1=\frac{1}{2}(\chi_{nn}-\chi_{np})$ is the isovector
strength constant. Usually one takes $\chi_1=\alpha\chi_0$,
$\alpha$ being a fitting parameter.
For the isoscalar strength constant $\chi_0$ we will take
the self consistent value
\begin{equation}
\chi_0=-\frac{3m\bar\omega^2}{Q_{00}}.
\label{self}
\end{equation}
Following ref. \cite{BaSc2}
we take $\alpha=-2$, i.e. a repulsive interaction with magnitude
twice as large as the isoscalar one.
Deriving (\ref{Ivect}) we used the self consistent expressions for
the oscillator frequencies \cite{BaSc2}
\begin{equation}
\omega_x^2=\omega_y^2=\bar\omega^2(1+\frac{4}{3}\delta),\quad
\omega_z^2=\bar\omega^2(1-\frac{2}{3}\delta),\quad
\bar\omega^2=\omega^2/(1+\frac{2}{3}\delta)
\label{omegaxz}
\end{equation}
and the standard \cite{BM} definition of the deformation parameter
$\delta=3Q_{20}/(4Q_{00})$, where
$Q_{00}=2\int\! d(\bp,\br) (x^2+y^2+z^2)f^0(\br,\bp)
=2\langle x^2\rangle+\langle z^2\rangle$
and $Q_{20}=2\langle z^2\rangle-2\langle x^2\rangle$ are
monopole and quadrupole moments of nucleus, respectively.

This set of equations has two integrals of motion :
\begin{eqnarray}
 \frac{i\hbar}{2m\Delta}\underline{\tilde P}
+[\frac{\hbar^2\chi_1 k_0}{4m\Delta}
-2m\bar\omega^2(1-\alpha)(1+\frac{\delta}{3})]
\frac{i\hbar}{4\Delta+2\chi_1 k_4}\underline{\tilde Q}-\underline{L}=\mathit{const}
\label{Int2iv}
\end{eqnarray}
and
\begin{equation}
 \underline{I_y}
+2m\bar\omega^2\delta(1-\alpha)
\frac{i\hbar}{4\Delta+2\chi_1 k_4}
\underline{\tilde Q}
=\mathit{const}.
\label{Int1iv}
\end{equation}
Obviously these constants should be equal zero. By definition the
variable $\underline{\tilde Q}$ is purely imaginary because $\kappa_-$
is the imaginary part of the pairing field $\kappa$. Therefore
Eq.~(\ref{Int1iv}) implies that the relative angular momentum
$\underline{I_y}$ oscillates in phase with the relative quadrupole
moment $\underline{\tilde Q}$ of the imaginary part of the pairing
field $\kappa$.

Analogously one can interpret Eq.~(\ref{Int2iv}) saying, that
the variable $\underline{L}$ oscillates out of phase with the linear
combination of two variables $\underline{\tilde Q}$ and
$\underline{\tilde P}$ which describe the quadrupole deformation of
the pairing field in coordinate and momentum spaces respectively.

\subsection{Eigenfrequencies}

Imposing the time evolution via $e^{i\Omega t}$ for all
variables one transforms Eqs.~(\ref{Ivect}) into
a set of algebraic equations, whose determinant gives the
eigenfrequencies of the system. We have
\begin{equation}
Det=(\E^2-2\chi_1 k_4\Delta)(\E^2-2\epsilon^2)
-2\epsilon^2(1-\alpha)\E^2-\chi_1\Delta k_0\hbar^4/m^2
+4\hbar^4\bar\omega^4(1-\alpha)\delta^2=0,
\label{charac1}
\end{equation}
where $\E^2=E^2-4\Delta^2$, $E=\hbar\Omega$,
$\epsilon^2=\hbar^2\bar\omega^2(1+\frac{\delta}{3})$.

In the case of $\Delta=0$ this equation is reduced to the known
\cite{BaSc2,Hamam} equation for the scissors mode. In the case
$\Delta\neq 0$ there are two solutions:
\begin{eqnarray}
E^2_{\pm}&=&4\Delta^2+
[\epsilon^2(2-\alpha)+\chi_1 k_4\Delta]
\nonumber\\
&&\pm\sqrt{[\epsilon^2(2-\alpha)+\chi_1 k_4\Delta]^2
-\chi_1\Delta [4k_4\epsilon^2-k_0\hbar^4/m^2]
-4\hbar^4\bar\omega^4(1-\alpha)\delta^2}.
\label{E2pm}
\end{eqnarray}
They describe the energies of the isovector GQR ($E_+$) and of the
scissors mode ($E_-$).

It is worth noting that contrary to the case without pairing
\cite{BaSc2} the energy of the scissors mode does not go to zero for
deformation $\delta=0.$ However this does not mean any contradiction
with the known quantum mechanical statement that the rotation of
spherical nuclei is impossible. It is easy to see from (\ref{Int1iv})
that the relative angular momentum $\underline{I_y}$ is conserved in
this case, $\underline{I_y}=\mathit{const}$, so the nature of this
mode of a spherical nucleus has nothing in common with the vibration of
angular momentum. The calculation of transition probabilities (see
below) shows that this mode can be excited by an electric field and it
is not excited by a magnetic field. Our estimation gives for the
energy of this mode the value about 4 MeV, that agrees very well with
the result of M. Matsuo et al \cite{Matsuo}, who studied the isovector
quadrupole response of $^{158}$Sn in the framework of QRPA with Skyrme
forces and found the proper resonance at $\sim$4.2 MeV.

It is known \cite{BaSc2,Zaw} that without pairing the scissors mode
has a non zero value of an energy only due to the Fermi Surface
Deformation (FSD). Let us investigate the role of FSD in the case with
pairing. Omitting in (\ref{Ivect}) the variable $\underline{P}$
responsible for FSD and its dynamical equation we obtain the following
characteristic equation
$$
E^2[E^2-4\Delta^2-2\alpha\chi_0 k_4\Delta-2\epsilon^2(1-\alpha)]=0.
$$

Two solutions $E_{sc}^2=0$ and
$E_{iv}^2=4\Delta^2+2\alpha\chi_0 k_4\Delta+2\epsilon^2(1-\alpha)$
reproduce the situation observed without pairing: the role of FSD is
crucial for the scissors mode and is not very important for IVGQR.

\subsection{Transition probabilities}

The transition probabilities are calculated with the help of the
linear response theory. The detailed description of its use within the
framework of WFM method can be found in \cite{BaSc2}, so we only
present the final results.

Electric quadrupole excitations are described by the operator
\begin{equation}
\label{Oelec}
\hat F=\hat F_{2\mu}^{\rm p}=\sum_{s=1}^Z\hat f_{2\mu}(s), \quad
\hat f_{2\mu}=e\,r^2Y_{2\mu}.
\end{equation}
The transition probabilities are
\begin{eqnarray}
\label{E2nu}
B(E2)_{\nu}=2|<\nu|\hat F_{21}^{\rm p}|0>|^2
=\frac{e^2\hbar^2}{m}\frac{5}{16\pi}Q_{00}
\frac{(1+\delta/3)(E_{\nu}^2-4\Delta^2)-2(\hbar\bar\omega\delta)^2}
{E_{\nu}[E^2_{\nu}-4\Delta^2-\epsilon^2(2-\alpha)-\chi_1 k_4\Delta]}.
\end{eqnarray}

Magnetic dipole excitations are described by the operator
\begin{equation}
\label{Omagn}
\hat F=\hat F_{1\mu}^{\rm p}=\sum_{s=1}^Z\hat f_{1\mu}(s), \quad
\hat f_{1\mu}=-i\nabla
(rY_{1\mu})\cdot[\br\times\nabla]\mu_N,\quad
\mu_N=\frac{e\hbar}{2mc}.
\end{equation}
 For transition probabilities we have
\begin{eqnarray}
\label{scimat}
B(M1)_{\nu}=2|<\nu|\hat F_{11}^{\rm p}|0>|^2=
\frac{1-\alpha}{8\pi}m\bar\omega^2
Q_{00}\delta^2\frac{E_{\nu}^2-4\Delta^2-2\epsilon^2}
{E_{\nu}[E^2_{\nu}-4\Delta^2-\epsilon^2(2-\alpha)-\chi_1 k_4\Delta]}
\,\mu_N^2.
\end{eqnarray}

Multiplying B(M1) factors of both states by the proper energies and
summing we find the following formula for the energy weighted sum rule
\begin{eqnarray}
\label{Msum}
E_{sc}B(M1)_{sc}+E_{iv}B(M1)_{iv}=(1-\alpha)
\frac{m\bar\omega^2}{4\pi}
Q_{00}\delta^2\mu_N^2.
\end{eqnarray}
This expression coincides exactly with the respective
sum rule calculated in \cite{BaSc2} without pairing. This means that
there is no contribution to the sum rule which comes from pairing.
This result can be explained by our approximation $\Delta=\mathit{const}$.

It is now a good place to discuss the deformation dependence of the
energies and transition probabilities. First we
recall the corresponding formulae without pairing:
\begin{eqnarray}
(E^0_{iv})^2=4\hbar^2\bar\omega^2
\left(1+\frac{\delta}{3}+\sqrt{(1+\frac{\delta}{3})^2-
 \frac{3}{4}\delta^2}\,\right),
\nonumber\\
(E^0_{sc})^2=4\hbar^2\bar\omega^2
\left(1+\frac{\delta}{3}-\sqrt{(1+\frac{\delta}{3})^2-
\frac{3}{4}\delta^2}\,\right),
\nonumber\\
B(M1)^0_{\nu}=
\frac{3}{8\pi}m\bar\omega^2
Q_{00}\delta^2\frac{E_{\nu}^2-2\epsilon^2}
{E_{\nu}(E^2_{\nu}-4\epsilon^2)}\,\mu_N^2,
\label{M1sc}
\end{eqnarray}
where the superscript ``0''means the absence of pairing and we
assumed $\alpha=-2$. The scissors mode energy
is proportional to $\delta$, that becomes evident after
expanding the square root:
\begin{equation}
\label{Esc0}
E^0_{sc}=\hbar\bar\omega\delta\sqrt{\frac{3}{2\delta_3}}
\left(1+\frac{3}{16}\frac{\delta^2}{\delta_3^2}
+\frac{9}{128}\frac{\delta^4}{\delta_3^4}+...\right),
\end{equation}
where $\delta_3=1+\delta/3.$

At a first superficial glance, the transition probability, as given
by formula (\ref{M1sc}), has the desired (experimentally observed)
quadratic deformation dependence.
However, due to the linear $\delta$-dependence of the factor $E_{sc}$
in the denominator, the resulting $\delta$-dependence of
$B(M1)_{sc}^0$ turns out to be linear, too. The situation is changed
radically when pairing is included. In this case the main contribution
to the scissors mode energy comes from the pairing interaction (the
term $4\Delta^2$ in (\ref{E2pm})),
$E_{sc}$ is not proportional to $\delta$ and the deformation
dependence of $B(M1)_{sc}$ becomes quadratic in excellent agreement
with QRPA calculations and experimental data
\cite{Zaw,Lo2000,Magnus,Garrido,Hamam,Sushkov,Macfar,Hilt86}.

The deformation dependence of $B(M1)_{iv}$ is quadratic in $\delta$,
even without pairing, because the energy $E_{iv}$ is not proportional
to $\delta$ and depends only weakly on it. The pairing does not change
this picture.

\section{Numerical results and discussion}

\begin{figure}
\begin{center}
\epsfig{file=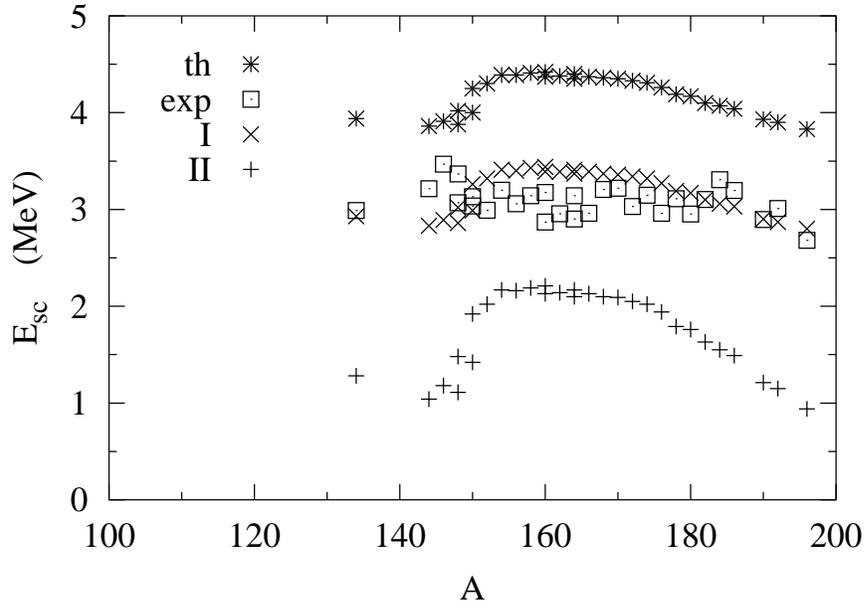,width=12cm}
\end{center}
\caption{\small Scissors mode energies as a function of the mass number $A$
for the nuclei listed in Table 1. The meaning of the symbols is
explained in the caption of Table 1.}
\label{figure1}
\end{figure}

We have reproduced all experimentally observed qualitative features of
the scissors mode. We understand that our model is too simplified to
describe also precise quantitative experimental characteristics.
Nevertheless we
performed the calculations of energies and $B(M1)$ factors to get at
least an idea on the order of magnitude of the discrepancy with
experimental data. The results of calculations for most nuclei, where
this mode is observed, are presented in Table \ref{table1} and in
Figures \ref{figure1}--\ref{figure4}. Formulae (\ref{E2pm}) and
(\ref{scimat}) were used with the following values of parameters:
$\alpha=-2$, $Q_{00}=A\frac{3}{5}R^2$, $R=r_0A^{1/3}$, $r_0=1.2$ fm,
$\bar\omega^2=\omega_0^2/
[(1+\frac{4}{3}\delta)^{2/3}(1-\frac{2}{3}\delta)^{1/3}]$,
$\hbar\omega_0=41/A^{1/3}$ MeV, $\hbar^2/m=41.803$ MeV fm$^2$. The gap
$\Delta$ as well as the integrals $k_0$, $k_4$ were calculated by two
methods: the semiclassical one, which is summarized in Appendix C, and
the microscopical one described in Appendix D.

Let us analyse at first the results obtained with semiclassical values
of $\Delta$, $k_0$, $k_4$ (columns $th$ and $I$ of Table 1). This
method gives $\Delta \approx 1.32$ MeV for all the nuclei
considered here. The values of $k_0, k_4$ vary smoothly from $k_0=37.7$,
$k_4=4243.7$ fm$^4$ for A=134 to $k_0=54.7$, $k_4=9946.3$ fm$^4$ for
A=196. Analysis of Table 1 shows, that overall agreement of theoretical
results with experimental data is reasonable. It is, of course, not
perfect, but the influence of pairing, especially on $B(M1)$
\begin{table}
\caption{\small Scissors mode energies $E_{sc}$ (in MeV) and transition
probailities $B(M1)_{sc}$ (in units of $\mu_N^2$);
$exp$: experimental values, $th1$ and $th$: full theory with $k_0, k_4$
calculated by microscopical and semiclassical methods respectively, I:
$\Delta\neq 0$ but $k_0=k_4=0$, II: theory without pairing
(i.e. $\Delta=k_0=k_4=0$). The experimental values of $E_{sc}$, $\delta$
and $B(M1)$ are from Ref. \cite{Pietr} and references therein.
$E1$ is explained in section 5.}
\label{table1}
\vspace{-0.2cm}
\begin{center}
\begin{tabular}{|c|c|c||c|c|c|c|c||c|c|c|c|c|}
\hline
   & & &
\multicolumn{5}{|c||}{ $E_{sc}$}    &
\multicolumn{5}{|c|}{ $B(M1)_{sc}$}    \\
\cline{4-13}
 Nuclei & $\delta$ & $E_1$ & $exp$ & $th1$ & $th$ & I & II & $exp$ & $th1$ & $th$ &
 I & II  \\
\hline
 $^{134}$Ba & 0.14 & 2.56 & 2.99 & 3.67 & 3.94 & 2.93 & 1.28 & 0.56 & 1.22 & 1.16 & 1.72 & 3.90 \\[-2mm]
 $^{144}$Nd & 0.11 & 2.07 & 3.21 & 3.18 & 3.86 & 2.83 & 1.04 & 0.17 & 1.05 & 0.86 & 1.30 & 3.54 \\[-2mm]
 $^{146}$Nd & 0.13 & 2.35 & 3.47 & 3.58 & 3.91 & 2.89 & 1.18 & 0.72 & 1.21 & 1.13 & 1.69 & 4.14 \\[-2mm]
 $^{148}$Nd & 0.17 & 2.96 & 3.37 & 3.92 & 4.02 & 3.02 & 1.48 & 0.78 & 1.78 & 1.79 & 2.65 & 5.39 \\[-2mm]
 $^{150}$Nd & 0.22 & 3.83 & 3.04 & 4.24 & 4.25 & 3.26 & 1.92 & 1.61 & 2.81 & 2.94 & 4.28 & 7.26 \\[-2mm]
 $^{148}$Sm & 0.12 & 2.21 & 3.07 & 3.64 & 3.88 & 2.86 & 1.11 & 0.43 & 1.07 & 1.02 & 1.59 & 3.96 \\[-2mm]
 $^{150}$Sm & 0.16 & 2.83 & 3.13 & 3.97 & 4.00 & 2.99 & 1.42 & 0.92 & 1.63 & 1.68 & 2.50 & 5.26 \\[-2mm]
 $^{152}$Sm & 0.24 & 4.02 & 2.99 & 3.77 & 4.30 & 3.32 & 2.02 & 2.26 & 3.70 & 3.27 & 4.75 & 7.81 \\[-2mm]
 $^{154}$Sm & 0.26 & 4.32 & 3.20 & 3.77 & 4.39 & 3.41 & 2.17 & 2.18 & 4.42 & 3.79 & 5.50 & 8.65 \\[-2mm]
 $^{156}$Gd & 0.26 & 4.29 & 3.06 & 3.78 & 4.39 & 3.40 & 2.16 & 2.73 & 4.44 & 3.82 & 5.55 & 8.76 \\[-2mm]
 $^{158}$Gd & 0.26 & 4.36 & 3.14 & 3.73 & 4.41 & 3.43 & 2.19 & 3.39 & 4.77 & 4.01 & 5.84 & 9.12 \\[-2mm]
 $^{160}$Gd & 0.27 & 4.39 & 3.18 & 3.64 & 4.42 & 3.44 & 2.21 & 2.97 & 5.08 & 4.14 & 6.02 & 9.38 \\[-2mm]
 $^{160}$Dy & 0.26 & 4.24 & 2.87 & 3.50 & 4.37 & 3.39 & 2.13 & 2.42 & 4.91 & 3.89 & 5.68 & 9.03 \\[-2mm]
 $^{162}$Dy & 0.26 & 4.25 & 2.96 & 3.43 & 4.38 & 3.39 & 2.14 & 2.49 & 5.19 & 3.99 & 5.83 & 9.25 \\[-2mm]
 $^{164}$Dy & 0.26 & 4.31 & 3.14 & 3.36 & 4.40 & 3.41 & 2.17 & 3.18 & 5.59 & 4.17 & 6.09 & 9.59 \\[-2mm]
 $^{164}$Er & 0.25 & 4.17 & 2.90 & 3.85 & 4.35 & 3.37 & 2.10 & 1.45 & 4.40 & 3.94 & 5.77 & 9.26 \\[-2mm]
 $^{166}$Er & 0.26 & 4.23 & 2.96 & 3.78 & 4.37 & 3.39 & 2.13 & 2.67 & 4.73 & 4.12 & 6.03 & 9.59 \\[-2mm]
 $^{168}$Er & 0.26 & 4.18 & 3.21 & 3.70 & 4.36 & 3.37 & 2.10 & 2.82 & 4.85 & 4.11 & 6.04 & 9.67 \\[-2mm]
 $^{170}$Er & 0.26 & 4.15 & 3.22 & 3.61 & 4.35 & 3.36 & 2.09 & 2.63 & 5.02 & 4.14 & 6.08 & 9.79 \\[-2mm]
 $^{172}$Yb & 0.25 & 4.08 & 3.03 & 3.59 & 4.33 & 3.34 & 2.05 & 1.94 & 4.96 & 4.08 & 6.01 & 9.79 \\[-2mm]
 $^{174}$Yb & 0.25 & 4.02 & 3.15 & 3.48 & 4.31 & 3.32 & 2.02 & 2.70 & 5.10 & 4.05 & 5.98 & 9.82 \\[-2mm]
 $^{176}$Yb & 0.24 & 3.85 & 2.96 & 3.32 & 4.26 & 3.27 & 1.94 & 2.66 & 5.06 & 3.83 & 5.67 & 9.58 \\[-2mm]
 $^{178}$Hf & 0.22 & 3.57 & 3.11 & 3.65 & 4.19 & 3.19 & 1.79 & 2.04 & 3.90 & 3.40 & 5.06 & 9.00 \\[-2mm]
 $^{180}$Hf & 0.22 & 3.50 & 2.95 & 3.54 & 4.17 & 3.17 & 1.76 & 1.61 & 3.96 & 3.34 & 4.97 & 8.97 \\[-2mm]
 $^{182}$W  & 0.20 & 3.25 & 3.10 & 3.40 & 4.10 & 3.10 & 1.63 & 1.65 & 3.63 & 2.96 & 4.44 & 8.43 \\[-2mm]
 $^{184}$W  & 0.19 & 3.09 & 3.31 & 3.36 & 4.07 & 3.06 & 1.55 & 1.12 & 3.37 & 2.74 & 4.12 & 8.14 \\[-2mm]
 $^{186}$W  & 0.18 & 2.97 & 3.20 & 3.32 & 4.04 & 3.03 & 1.49 & 0.82 & 3.22 & 2.60 & 3.91 & 7.95 \\[-2mm]
 $^{190}$Os & 0.15 & 2.42 & 2.90 & 3.35 & 3.93 & 2.90 & 1.21 & 0.98 & 2.15 & 1.82 & 2.77 & 6.64 \\[-2mm]
 $^{192}$Os & 0.14 & 2.29 & 3.01 & 3.20 & 3.90 & 2.87 & 1.15 & 1.04 & 2.06 & 1.66 & 2.54 & 6.37 \\[-2mm]
 $^{196}$Pt & 0.11 & 1.87 & 2.68 & 3.00 & 3.83 & 2.80 & 0.94 & 0.70 & 1.52 & 1.16 & 1.79 & 5.35 \\
\hline
\end{tabular}
\end{center}
\end{table}
values,
is impressive. As it is seen, without pairing the calculated energies
(column II) are 1 -- 2 MeV (1.5 -- 2 times) smaller, than $E_{exp}$,
and $B(M1)$ factors (column II) exceed experimental values 3 -- 7
times. Taking into account $\Delta$ (without $k_4, k_0$) changes the
results (columns I) drastically: the differences between calculated
and experimental energies are reduced to 5 -- 10$\%$ and calculated
transitions probabilities are reduced by a factor of 1.5 --
2. Inclusion of $k_4$ (columns $th$) reduces the transition
probabilities again by a factor of 1.5, improving the agreement
appreciably, and increases the energies by $\sim$1 MeV deteriorating
slightly the agreement with experimental data. The influence of $k_0$
is negligibly small, being of the order of $\sim 1\%$.  Finally we
obtain that $E_{th}$ exceeds $E_{exp}$ by $\sim$1 -- 1.3 MeV and that
$B(M1)_{th}$ exceeds $B(M1)_{exp}$ approximately by a factor of 1.5 -- 2.

Let us analyse now the results obtained with microscopical values of
$\Delta$, $k_0$, $k_4$ (columns $th1$ of Table 1). Microscopical gaps
for neutrons and protons vary in the range 0.8 -- 1.0 MeV and 1.0 -- 1.2
MeV respectively, being appreciably smaller than the semiclassical
ones. The values of $k_4$ vary in the limits 4888 -- 8936 fm$^4$, being
in the reasonable agreement with semiclassical values (in general they
are smaller by $\sim$ 10 -- 20 $\%$). Generally they increase with the
atomic number A, however, due to shell effects it happens not
monotonically, as in the semiclassical case, but quite irregularly.
The integrals $k_0$ vary in the range 56 -- 78. They are very sensitive
to the detailes of level schemes -- that is why they differ
substantially (sometimes 60 -- 70$\%$) from the semiclassical
values, being nevertheless of the same order of magnitude.

The analysis of final results (Table 1) reveals the following trend:
the scissors mode energies
decrease (in comparison with those of column $th$) by $\sim$ 0.5 -- 0.7
MeV, that improves the agreement with experimental data, and B(M1)
factors increase by $\sim$ 10 -- 20 $\%$, that leads to a somewhat
worse agreement with experimental data.

\begin{figure}
\begin{center}
\epsfig{file=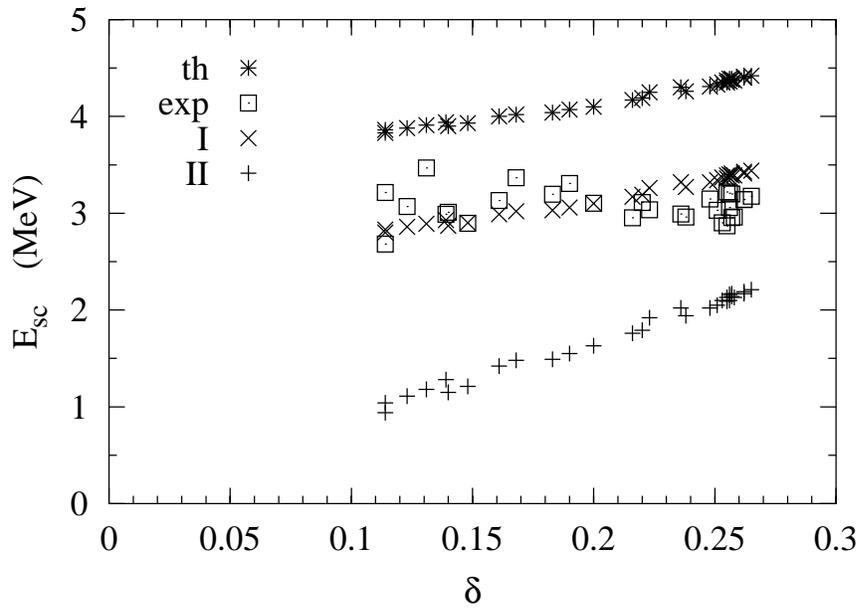,width=12cm}
\end{center}
\caption{\small Scissors mode energies as a function of the deformation $\delta$
for the nuclei listed in Table 1. The meaning of the symbols is
explained in the caption of Table 1.}
\end{figure}
\begin{figure}
\begin{center}
\epsfig{file=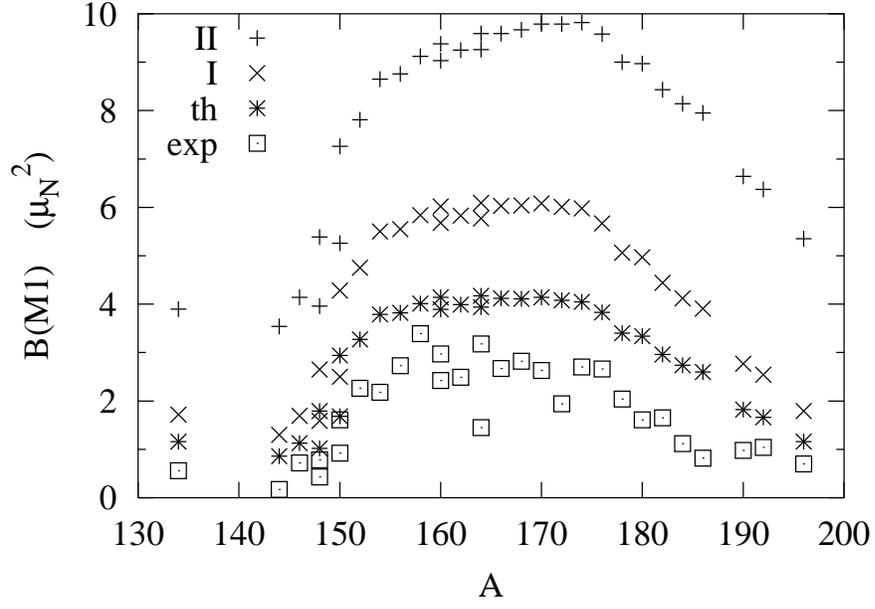,width=12cm}
\end{center}
\caption{\small Scissors mode transition probabilities $B(M1)$ as a function
of the mass number $A$ for the nuclei listed in Table 1. The meaning
of the symbols is explained in the caption of Table 1.}
\end{figure}
\begin{figure}
\begin{center}
\epsfig{file=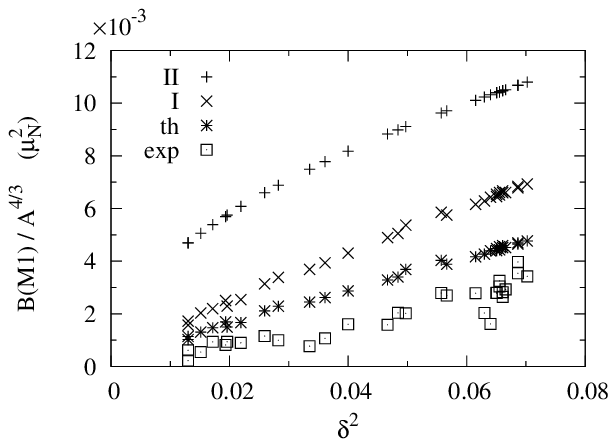,width=12cm}
\end{center}
\caption{\small Scissors mode transition probabilities, normalized by a
factor of $A^{-4/3}$, as a function of the deformation $\delta$ for the
nuclei listed in Table 1. The meaning of the symbols is explained in
the caption of Table 1.}
\label{figure4}
\end{figure}

What can be done to improve these results? The first step is obvious --
it is necessary to get rid off approximations enumerated at the
beginning of section 3, especially of the most crude one:
$\Delta^{\rm p}=\Delta^{\rm n}$. As a result, it will be necessary to solve
the coupled isoscalar and isovector sets of equations. The next possible
step is to perform self consistent calculation with a more or
less realistic interaction and taking into account the $\br$-dependence of
$\Delta$.

Another point, which should be clarified, is the role of the
spin-orbit interaction. It is known \cite{Zaw}, that experimentally
observed low lying magnetic dipole strength consists of two separated
parts: orbital excitations in an energy interval $\sim$2 -- 4 MeV and
the spin-flip resonance ranging from 5 to 10 MeV excitation energy.
So, for the full description of the scissors mode dynamics it would be
necessary to consider also the spin degrees of freedom. One may expect
that the orbital part of the M1 strength (scissors mode) will be
pushed down by the spin-orbit interaction, in agreement with
experimental data.

\section{Currents}

According \cite{BaSc2} the components of infinitesimal displacements in the
plane $y=0$ are given by
\begin{equation}
\xi_x(t)=\sqrt2B\underline Q(t)z,
\quad \xi_z(t)=\sqrt2A\underline Q(t)x
\label{disp}
\end{equation}
with
\begin{eqnarray}
A=\frac{3}{\sqrt2}[1-2\frac{\bar\omega^2}{\Omega^2}(1-\alpha)\delta]
/[Q_{00}(1-\frac{2}{3}\delta)],
\nonumber\\
B=\frac{3}{\sqrt2}[1+2\frac{\bar\omega^2}{\Omega^2}(1-\alpha)\delta]
/[Q_{00}(1+\frac{4}{3}\delta)].
\label{AiB}
\end{eqnarray}
 It is useful to write these displacement as the superposition of a
rotational component with the coefficient $a$ and an irrotational
one with the coefficient $b$
\begin{equation}
\vec\xi=a\vec e_y\times\vec r+b\nabla(xz)=a(z,0,-x)+b(z,0,x)\quad
\longrightarrow\quad
\xi_x=(b+a)z, \quad \xi_z=(b-a)x.
\label{disp2}
\end{equation}
Comparison of  (\ref{disp}) with  (\ref{disp2}) gives
$$
b+a=\sqrt2B\underline Q, \quad b-a=\sqrt2A\underline Q \quad
\longrightarrow \quad b=(B+A)\underline Q/\sqrt2,\quad
a=(B-A)\underline Q/\sqrt2.
$$
Using here expressions  (\ref{AiB}) one finds
\begin{equation}
b=\gamma(\delta_3-\delta^2/g), \quad a=\gamma\delta(1-\delta_3/g),
\label{aib}
\end{equation}
where $g=E^2/(6\hbar^2\bar\omega^2)$, $\delta_3=1+\delta/3$ and
$\gamma=3\underline Q/[O_{00}(1-\frac{2}{3}\delta)(1+\frac{4}{3}\delta)]$.
The coefficients $a, b$ are shown schematically as the functions of
$g$ on the figure 5. In spite of the simple behaviour of two curves,
they describe rather interesting phenomena.
There are two critical points: $g_1=\delta^2/\delta_3$, where $b=0$,
and $g_2=\delta_3$, where $a=0$. They divide $g$ axes into
three regions.
\begin{figure}
\begin{center}
\epsfig{file=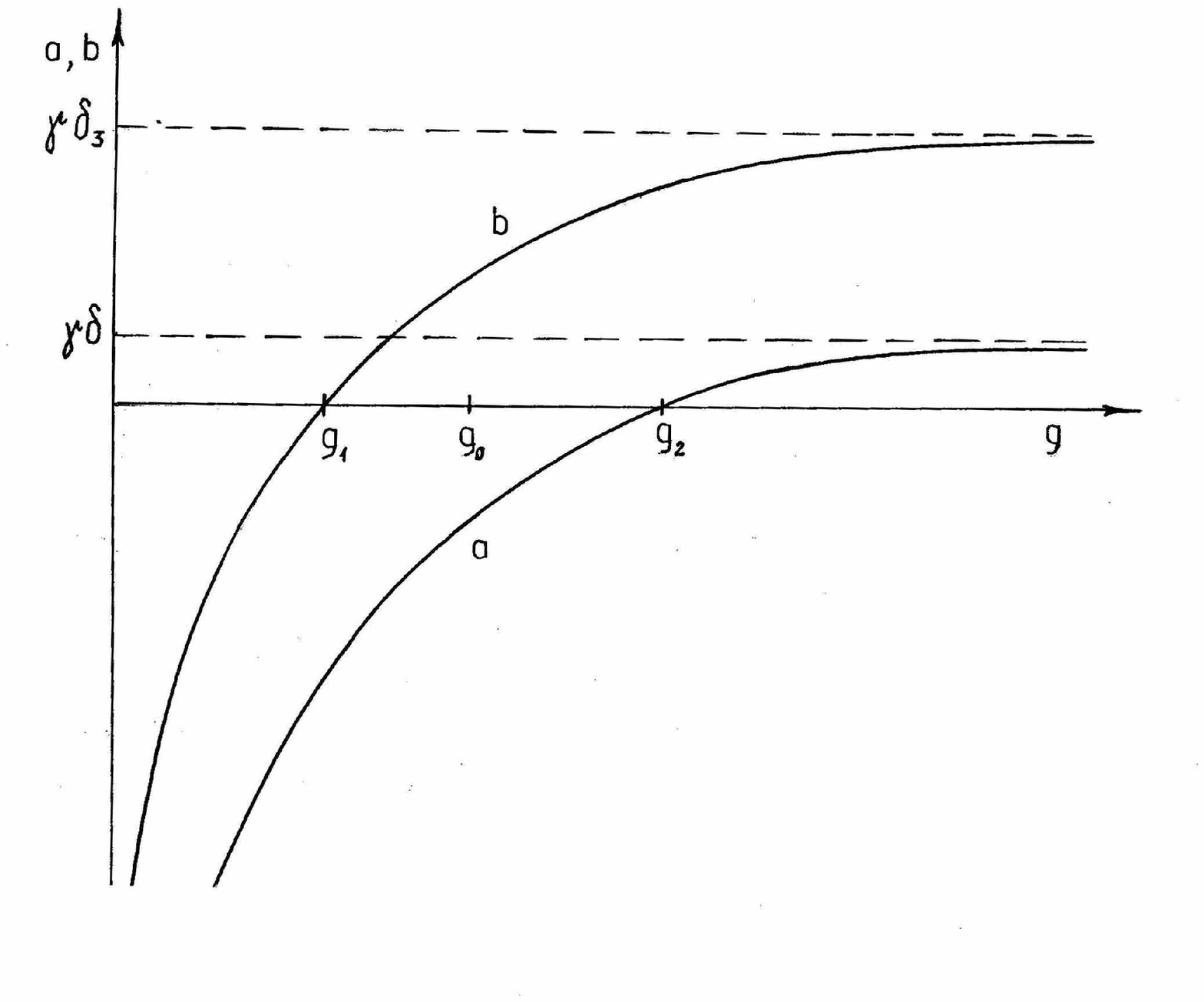,width=12cm}
\end{center}
\caption{\small Amplitudes $a$ and $b$ (in arbitrary units) of the
rotational and irrotational current components as the functions of
energy ($g=E^2/(6\hbar^2\bar\omega^2)$).}
\label{figure5}
\end{figure}

In the region $0<g<g_1$ the rotational component of the current
dominates. The ratio $b/a$ is changed from $\delta/\delta_3$ at $g=0$
to zero at $g=g_1$. At this point the motion is pure rotational! The
energy corresponding to this point is
$E_1=\hbar\bar\omega\delta\sqrt{6/\delta_3}.$ Its value for
various nuclei is shown in the Table 1. It is interesting that all
experimental values of $E_{sc}$ are disposed  just in the vicinity of
$E_1$, the
energies of nuclei with the small deformation ($\delta \le .19$)
being disposed on the right hand side of this point, while the
energies of nuclei with large deformations ($\delta \ge .2$) are
disposed on the left hand side.
The theoretical value of $E_{sc}$ obtained without pairing (see
formula  (\ref{Esc0})) is $E_{sc}^0\simeq \frac{1}{2}E_1$, i.e. it is
disposed far to the left of $E_1$ for all nuclei independently of their
deformations. Inclusion of pairing shifts the scissors mode energy
$E_{sc}$ strongly to the right so, that it becomes larger than the
critical value $E_1$, but quite close to it, especially
in the well deformed nuclei. One can note, that the
fact of the proximity of theoretical and, especially, experimental
values of $E_{sc}$ to the critical point $E_1$ gives some grounds for
the two rotors model of the scissors mode \cite{Lo2000}, where the
pure rotational motion of the nuclear matter is supposed.

In the region $g_1 < g < g_2$ the rotational and irrotational
components of motion compete on an equal footing -- the dominance of
the rotational motion at the point $g=g_1$ is gradually replaced by
the dominance of the irrotational motion at the point $g=g_2$, the
strengths of both components being equal ( i.e. $|a|=|b|$ ) at the
point $g_0=\delta$. The energy corresponding to the point $g_2$ is
$E_2=\sqrt{6\delta_3}\hbar\bar\omega
=\sqrt{3\delta_3}\sqrt2\hbar\bar\omega$. The energy centroid of IVGQR
(without pairing) is $E_{iv}=2\sqrt2\hbar\bar\omega$. For normally
deformed nuclei ($\delta<1$) the factor $\sqrt{3\delta_3}$ is smaller
than 2. Therefore $E_{iv}$ is disposed on the right hand side of the
point $g_2$, i. e. in the region $g_2<g<\infty$, where the
irrotational component of the current dominates.
The inclusion of pairing shifts $E_{iv}$ more to the right.
 With $g$ increasing the coefficients
$a$ and $b$ grow gradually aspiring to their asymptotical values
$\gamma\delta$ and $\gamma\delta_3$ respectively.

One more detail of the interrelation of the two components of the
current. Let us observe the motion of the ends (tips) of the
scissors. Their displacements are determined essentially by the value
of $\xi_x=(b+a)z$. In the region $0<g<g_1$ coefficients $a$ and $b$
have the opposite signs.
This means that two components of current move out of phase --
the shear (irrotational) component bends the bodies of the scissors
trying to resist to their rotation. This process is described very
well and illustrated by the figures 1, 2 in the paper by R. Hilton
\cite{Hilt92}. In the region $g_1<g<g_2$ coefficients $a$ and $b$
have the same sign.
Therefore, both components move in phase -- now the shear
component bends the bodies of the scissors reinforcing their rotation.
In the region $g_2<g<\infty$ coefficients $a$ and $b$ again have
opposite signs and
the situation returns to the case of the first region. However, now
the shear component dominates, hence it is better to say that the
rotation trys to resist to the shear.

\section{Concluding remarks}

The low energy magnetic dipole strength
produced in all QRPA calculations (even in the schematic model HO+QQ)
 is always distributed over several
states in the region $0<E<4$ MeV, because each $\alpha\beta$ pair
contributing to $B(M1)_{sc}$ occurs at a different energy
$E_{\alpha\beta}=E_{\alpha}+E_{\beta}$ (see Introduction).
The same picture is observed experimentally. These facts were in
sharp contradiction with the anticipated single peak, predicted in
early papers \cite{Suzuki,Lipp,Bes} and caused ``controversy
about the collectivity and the correct interpretation - in the sense
of classical motion - of the orbital magnetic dipole modes
\cite{Macfar}''. It became clear that one needs some kind of a bridge
between the classical and quantum mechanical approaches. For the
description of this phenomenon ``a formalism is required in which the
interplay between the collective and single particle aspects of the
system are adequately treated \cite{Hilt86}''.

Some authors ``have tried to complement RPA calculations with
realistic forces by classical and semiclassical methods which are
flexible enough also to admit other solutions than the one anticipated
\cite{Macfar}''. This way however does not give a satisfactory solution
of the problem, because usually the phenomenological models have not
the direct connection with quantum mechanical ones. The most popular
way to extract the underlying physical nature of the state from
microscopic (RPA or QRPA) calculations is the calculation of overlaps
\cite{Zaw}. However, the overlaps usually do not give the full
information about the various properties of the studied phenomenon. As
a matter of fact, this procedure answers only the question: ``What
part of the mode strength is excited by this operator (with which the
overlap is calculated)?''  Usually one calculates the overlap with the
``synthetic scissors'' \cite{Lo2000,Hamam}. It is known, however, that
due to the coupling with the IVGQR \cite{Lipp}, the low-energy magnetic
dipole mode contains rather big isovector plus isoscalar admixtures of
two orthogonal shears, as shown in \cite{Hilt86} by calculating
the overlap with the proper operator. Nevertheless, this overlap does
not exhaust all strength. What other operators are necessary to
clarify completely the nature of the considered mode?

We think that the combination of two complementary methods, namely RPA
(or QRPA) and WFM method, can be very useful. Based on the same
approach (time-dependent HF or HFB together with a small amplitude
approximation), they allow one, starting from the same Hamiltonian
with the same forces, to obtain either (by RPA calculations) the
refined microscopic structure or (applying WFM method) the crude
``macroscopic'' picture (which reminds very much the results of
semiclassical approaches) of the same phenomenon. This interrelation
of RPA and WFM was investigated in \cite{BaSc3}. In particular, the
identity of both methods in the case of a schematic model was
demonstrated there. With the help of the WFM method, taking into account
moments of higher and higher rank, one can produce a more and more
detailed description of the phenomenon, achieving (at least in
principle) the maximally fragmented picture given by the experiment
(and RPA).

One more remark is in order. Discussing the scissors mode energy and its
$\delta$-dependence, one has usually in mind the mean excitation
energy (centroid) defined as the center of gravity of the M1 strength
distributed among the low-lying $K^{\pi}=1^+$ states
$$
E_{sc}=\sum_iE_iB(M1)_i/\sum_iB(M1)_i.
$$
The sums are evaluated in the energy intervals around 3 MeV. The
evaluation interval is not strictly determined. For example, Hamamoto
and Nazarewicz \cite{Hamam} took $0<E<10$ MeV in their calculations of
superdeformed nuclei. Applying the WFM method, which produces
centroids of resonances, we avoid such problems.

In conclusion, the WFM method is generalized to take into account pair
correlations and is used to calculate energies and transitions
probabilities of the scissors mode. Excellent qualitative and
reasonable quantitative agreement with experimental data is obtained.
In addition the interrelation of microscopic and semiclassical
features of the scissors mode is clarified.

\section*{Acknowledgements}

Fruitful discussions with A. A. Dzhioev are gratefully acknowledged.
\section*{Appendix A}

We remind that each ingredient of Eqs.~(\ref{HFB}) is a matrix
in coordinates and spin projections. For example,
$\hat\kappa\equiv\hat\kappa_{\br,\sigma;\br',\sigma'}$. The spin structures of
the normal and abnormal density are explicitly
$$
\hat\rho_{\br,\sigma;\br',\sigma'}={\hat\rho_{\br,\br'}
\quad 0\quad\choose \quad 0\quad
\hat\rho_{\br,\br'}},\quad
\hat\kappa_{\br,\sigma;\br',\sigma'}=
{\quad 0\qquad \hat\kappa_{\br,\br'}
\choose -\hat\kappa_{\br,\br'}\quad 0\quad},
$$ or
$$
\hat\rho_{\br,\uparrow;\br',\uparrow}=
\hat\rho_{\br,\downarrow;\br',\downarrow},\quad
\hat\kappa_{\br,\uparrow;\br',\downarrow}=
-\hat\kappa_{\br,\downarrow;\br',\uparrow}.
$$
With the help of the standard expression for the product of two
matrices
$$(AB)_{\br,\sigma;\br',\sigma'}=\int d^3r''\sum_{\sigma''}
A_{\br,\sigma;\br'',\sigma''}B_{\br'',\sigma'';\br',\sigma'}=
\int d^3r''[A_{\br,\sigma;\br'',\uparrow}
B_{\br'',\uparrow;\br',\sigma'}
+A_{\br\sigma;\br'',\downarrow}
B_{\br'',\downarrow;\br'\sigma'}]$$
we find
$$(\hat h\hat \rho)_{\br,\uparrow;\br',\uparrow}=
\int d^3r''\hat h_{\br,\uparrow;\br'',\uparrow}
\hat\rho_{\br'',\uparrow;\br',\uparrow},\quad
(\hat h\hat\rho)_{\br,\uparrow;\br',\downarrow}=
(\hat h\hat\rho)_{\br,\downarrow;\br',\uparrow}=0,$$
$$(\hat h\hat\rho)_{\br,\downarrow;\br',\downarrow}=
\int d^3r''\hat h_{\br,\downarrow;\br'',\downarrow}
\hat\rho_{\br'',\downarrow;\br',\downarrow},$$
$$(\hat\Delta\hat\kappa)_{\br,\uparrow;\br',\uparrow}=
\int d^3r''\hat\Delta_{\br,\uparrow;\br'',\downarrow}
\hat\kappa_{\br'',\downarrow;\br',\uparrow},\quad
(\hat\Delta\hat\kappa)_{\br,\uparrow;\br',\downarrow}=
(\hat\Delta\hat\kappa)_{\br,\downarrow;\br',\uparrow}=0,$$
$$(\hat\Delta\hat\kappa)_{\br,\downarrow;\br',\downarrow}=
\int d^3r''\hat\Delta_{\br,\downarrow;\br'',\uparrow}
\hat\kappa_{\br'',\uparrow;\br',\downarrow}.$$
As an example we write out the first
equation of (\ref{HFB}) for $\sigma=\uparrow$ in detail:
\begin{eqnarray}
i\hbar\dot{\hat\rho}_{\br,\uparrow;\br',\uparrow}=
\int d^3r''
\left[
\hat h_{\br,\uparrow;\br'',\uparrow}
\hat\rho_{\br'',\uparrow;\br',\uparrow}
-\hat\rho_{\br,\uparrow;\br'',\uparrow}
\hat h_{\br'',\uparrow;\br',\uparrow}
\right.
\nonumber\\
\hspace{2cm}\left.
-\hat\Delta_{\br,\uparrow;\br'',\downarrow}
\hat\kappa^{\dagger}_{\br'',\downarrow;\br',\uparrow}
+\hat\kappa_{\br,\uparrow;\br'',\downarrow}
\hat\Delta^{\dagger}_{\br'',\downarrow;\br',\uparrow}
\right].
\label{dotrho}
\end{eqnarray}

The Wigner Transform (WT) of the single particle operator matrix
$\hat F_{\br_1,\sigma;\br_2,\sigma'}$ is defined as
$$[\hat F_{\br_1,\sigma;\br_2,\sigma'}]_{WT}\equiv
F_{\sigma,\sigma'}(\br,\bp)
=\int d^3s e^{-i\bp\cdot\bs/\hbar}
\hat F_{\br+\bs/2,\sigma;\br-\bs/2,\sigma'}$$
with $\br=(\br_1+\br_2)/2$ and $\bs=\br_1-\br_2.$
It is easy to derive a pair of useful relations. The first one is
$$F_{\sigma,\sigma'}^*(\br,\bp)
=\int d^3s e^{i\bp\cdot\bs/\hbar}
\hat F^*_{\br+\bs/2,\sigma;\br-\bs/2,\sigma'}
=\int d^3s e^{-i\bp\cdot\bs/\hbar}
\hat F^*_{\br-\bs/2,\sigma;\br+\bs/2,\sigma'}$$
$$=\int d^3s e^{-i\bp\cdot\bs/\hbar}
\hat F^{\dagger}_{\br+\bs/2,\sigma';\br-\bs/2,\sigma}=
[\hat F^{\dagger}_{\br_1,\sigma';\br_2,\sigma}]_{WT}$$
i.e. $[\hat F^{\dagger}_{\br_1,\sigma;\br_2,\sigma'}]_{WT}
=[\hat F_{\br_1,\sigma';\br_2,\sigma}]_{WT}^*
=F_{\sigma'\sigma}^*(\br,\bp).$
The second relation is
$$\bar F_{\sigma\sigma'}(\br,\bp)\equiv F_{\sigma\sigma'}(\br,-\bp)
=\int d^3s e^{i\bp\cdot\bs/\hbar}
\hat F_{\br+\bs/2,\sigma;\br-\bs/2,\sigma'}$$
$$=\int d^3s e^{-i\bp\cdot\bs/\hbar}
\hat F_{\br-\frac{\bs}{2},\sigma;\br+\frac{\bs}{2},\sigma'}
=\int d^3s e^{-i\bp\cdot\bs/\hbar}
[\hat F^{\dagger}_{\br+\bs/2,\sigma';\br-\bs/2,\sigma}]^*.$$
For the hermitian operators $\hat \rho$ and $\hat h$ this latter relation gives
$[\hat\rho^*_{\br_1,\sigma;\br_2,\sigma}]_{WT}
=\rho_{\sigma\sigma}(\br,-\bp)$ and
$[\hat h^*_{\br_1,\sigma;\br_2,\sigma}]_{WT}
=h_{\sigma\sigma}(\br,-\bp)$.

The Wigner transform of the product of two matrices $F$ and $G$ is
\begin{eqnarray}
[\hat F\hat G]_{WT}=F(\br,\bp)\exp\left(\frac{i\hbar}{2}
\stackrel{\leftrightarrow}{\Lambda}\right)G(\br,\bp),
\label{WigFG}
\end{eqnarray}
where the symbol
 $\stackrel{\leftrightarrow}{\Lambda}$
 stands for the Poisson bracket operator
$$
\stackrel{\leftrightarrow}{\Lambda}
=\sum_{i=1}^3\left(
\frac{\stackrel{\gets}{\partial} }{\partial r_i}\frac{
\stackrel{\to}{\partial} }{\partial p_i}
-\frac{\stackrel{\gets}{\partial} }{\partial p_i}\frac{
\stackrel{\to}{\partial} }{\partial r_i}\right).$$
For example the Wigner transform of Eq.~(\ref{dotrho})
up to linear order in $\hbar$ is
\begin{eqnarray}
i\hbar\dot f_{\uparrow,\uparrow}(\br,\bp)&=&
i\hbar\{h_{\uparrow,\uparrow}(\br,\bp),
f_{\uparrow,\uparrow}(\br,\bp)\}
\nonumber\\&&
-\Delta_{\uparrow,\downarrow}(\br,\bp)
\kappa^*_{\uparrow,\downarrow}(\br,\bp)
-\frac{i\hbar}{2}\{ \Delta_{\uparrow,\downarrow}(\br,\bp),
\kappa^*_{\uparrow,\downarrow}(\br,\bp)\}
\nonumber\\
&&+\kappa_{\uparrow,\downarrow}(\br,\bp)
\Delta^*_{\uparrow,\downarrow}(\br,\bp)
+\frac{i\hbar}{2}\{ \kappa_{\uparrow,\downarrow}(\br,\bp),
\Delta^*_{\uparrow,\downarrow}(\br,\bp)\},
\label{Wdotrho}
\end{eqnarray}
where $\di\{f,g\}\equiv f\stackrel{\leftrightarrow}{\Lambda}g
=\sum_{i=1}^3\left(
\frac{\partial f}{\partial r_i}\frac{\partial g}{\partial p_i}
-\frac{\partial f}{\partial p_i}\frac{\partial g}{\partial r_i}
\right)$ is the Poisson bracket of arbitrary functions $f(\br,\bp)$
and $g(\br,\bp)$; $h(\br,\bp)$, $f(\br,\bp)$, $\Delta(\br,\bp)$ and
$\kappa(\br,\bp)$ are Wigner transforms of
$h_{\br_1,\br_2}$, $\rho_{\br_1,\br_2}$, $\Delta_{\br_1,\br_2}$ and
$\kappa_{\br_1,\br_2}$ respectively. The functions
$h$ and $f$ are real, because the matrices $\hat h$ and $\hat \rho$ are
hermitian.
This example demonstrates in an obvious way that the dynamical equations
(\ref{HFB}) for the matrix elements
$\rho_{\br,\sigma;\br',\sigma}$,
$\rho^*_{\br,\sigma;\br',\sigma}$,
$\kappa_{\br,\pm\sigma;\br',\mp\sigma}$,
$\kappa^{\dagger}_{\br,\pm\sigma;\br',\mp\sigma}$,
with $\sigma=\uparrow$ and $\sigma=\downarrow$ are transformed
into eight dynamical equations for their Wigner transforms: 4 equations
for
$f_{\sigma,\sigma}(\br,\bp)$,
$\bar f_{\sigma,\sigma}(\br,\bp)$,
$\kappa_{\sigma,-\sigma}(\br,\bp)$,
$\kappa^*_{\sigma,-\sigma}(\br,\bp)$
with $\sigma=\uparrow$ and 4 equations with $\sigma=\downarrow$.
By definition $\bar f_{\sigma,\sigma}(\br,\bp)=
f_{\sigma,\sigma}(\br,-\bp)$.
In the absence of spin dependent forces both of these subsets
coincide and we can consider any one of them.

\section*{Appendix B}

Let us consider in detail the arbitrary term of equations (\ref{HFBA})
proportional to $\hbar^n$. Such terms appear after expanding the
exponent in formula (\ref{WigFG}). After integration with an arbitrary
function $A(\br,\bp)$ we have
\begin{eqnarray}
I_n&=&\int\! d(\bp,\br)A[f \stackrel{\leftrightarrow}{\Lambda}^n g]=
\int\! d(\bp,\br)\left(
A\left[\frac{\partial f}{\partial x_i}
\stackrel{\leftrightarrow}{\Lambda}^{n-1}
\frac{\partial g}{\partial p_i}\right]-
A\left[\frac{\partial f}{\partial p_i}
\stackrel{\leftrightarrow}{\Lambda}^{n-1}
\frac{\partial g}{\partial x_i}\right]\right)
\nonumber\\
&=&-\int\! d(\bp,\br)\left(
\frac{\partial A}{\partial p_i}\left[\frac{\partial f}{\partial x_i}
\stackrel{\leftrightarrow}{\Lambda}^{n-1} g\right]-
\frac{\partial A}{\partial x_i}\left[\frac{\partial f}{\partial p_i}
\stackrel{\leftrightarrow}{\Lambda}^{n-1} g\right]
\right)
\equiv A_p+A_x.
\label{arbitr}
\end{eqnarray}
Repeating the integration by parts we find
\begin{eqnarray}
A_p&=&-\int\! d(\bp,\br)
\frac{\partial A}{\partial p_i}
\left(
\left[
\frac{\partial^2 f}{\partial x_i\partial x_j}
\stackrel{\leftrightarrow}{\Lambda}^{n-2}
\frac{\partial g}{\partial p_j}\right]-
\left[
\frac{\partial^2 f}{\partial x_i\partial p_j}
\stackrel{\leftrightarrow}{\Lambda}^{n-2}
\frac{\partial g}{\partial x_j}\right]
\right)
\nonumber\\
&=&\int\! d(\bp,\br)\left(
\frac{\partial^2 A}{\partial p_i\partial p_j}
\left[\frac{\partial^2 f}{\partial x_i\partial x_j}
\stackrel{\leftrightarrow}{\Lambda}^{n-2} g\right]-
\frac{\partial^2 A}{\partial p_i\partial x_j}
\left[\frac{\partial^2 f}{\partial x_i\partial p_j}
\stackrel{\leftrightarrow}{\Lambda}^{n-2} g\right]
\right)
\equiv A_{pp}+A_{px}.
\nonumber\\
A_x&=&-\int\! d(\bp,\br)
\frac{\partial A}{\partial x_i}
\left(
\left[
\frac{\partial^2 f}{\partial p_i\partial x_j}
\stackrel{\leftrightarrow}{\Lambda}^{n-2}
\frac{\partial g}{\partial p_j}\right]-
\left[
\frac{\partial^2 f}{\partial p_i\partial p_j}
\stackrel{\leftrightarrow}{\Lambda}^{n-2}
\frac{\partial g}{\partial x_j}\right]
\right)
\nonumber\\
&=&\int\! d(\bp,\br)\left(
\frac{\partial^2 A}{\partial x_i\partial p_j}
\left[\frac{\partial^2 f}{\partial p_i\partial x_j}
\stackrel{\leftrightarrow}{\Lambda}^{n-2} g\right]-
\frac{\partial^2 A}{\partial x_i\partial x_j}
\left[\frac{\partial^2 f}{\partial p_i\partial p_j}
\stackrel{\leftrightarrow}{\Lambda}^{n-2} g\right]
\right)
\equiv A_{xp}+A_{xx}.
\label{arbitr2}
\end{eqnarray}
Repeating once again the integration by parts we get
\begin{eqnarray}
A_{pp}=\int\! d(\bp,\br)
\frac{\partial^2 A}{\partial p_i\partial p_j}
\left(
\left[
\frac{\partial^3 f}{\partial x_i\partial x_j\partial x_k}
\stackrel{\leftrightarrow}{\Lambda}^{n-3}
\frac{\partial g}{\partial p_k}\right]-
\left[
\frac{\partial^3 f}{\partial x_i\partial x_j\partial p_k}
\stackrel{\leftrightarrow}{\Lambda}^{n-3}
\frac{\partial g}{\partial x_k}\right]
\right)
\nonumber\\
=-\int\! d(\bp,\br)\left(
\frac{\partial^3 A}{\partial p_i\partial p_j\partial p_k}
\left[\frac{\partial^3 f}{\partial x_i\partial x_j\partial x_k}
\stackrel{\leftrightarrow}{\Lambda}^{n-3} g\right]-
\frac{\partial^3 A}{\partial p_i\partial p_j\partial x_k}
\left[\frac{\partial^3 f}{\partial x_i\partial x_j\partial p_k}
\stackrel{\leftrightarrow}{\Lambda}^{n-3} g\right]
\right).
\label{arbitr3}
\end{eqnarray}
It is easy to see from the structure of $A_{pp}$ (and $A_{px},
A_{xp}, A_{xx},$) that in the case, when $A(\br,\bp)$ is a polynomial
of an order $k$, all integrals $I_n$ with $n>k$ are equal to zero. In
our case $k=2$ and we find from the above formulae, that
$I_1=\int\! d(\bp,\br)A\{f,g\},\quad
I_2=\int\! d(\bp,\br)A\{\{f,g\}\},$ where $\{f,g\}$ is defined in
Appendix A and
$$\{\{f,g\}\}\equiv f(\br,\bp)
\stackrel{\leftrightarrow}{\Lambda}^2g(\br,\bp)=\sum_{i,j=1}^3\left(
\frac{\partial^2f}{\partial r_i\partial r_j}
\frac{\partial^2g}{\partial p_i\partial p_j}
-2\frac{\partial^2f}{\partial r_i\partial p_j}
\frac{\partial^2g}{\partial p_i\partial r_j}
+\frac{\partial^2f}{\partial p_i\partial p_j}
\frac{\partial^2g}{\partial r_i\partial r_j}\right).
$$

\section*{Appendix C: Thomas-Fermi approach to nuclear pairing}

Following Ref.\cite{VSFC}, we define the density matrix averaged on
the energy shell as

\begin{equation}
\hat{\rho}_E = \frac{1}{\tilde g(E)} \tilde \delta (E
-\hat{H})
= \frac{1}{\tilde g(E)} \sum_\nu \tilde \delta (E -
\varepsilon_\nu)
| \nu \rangle \langle \nu |.
\label{eq3} \end{equation}
which is a smooth function of $E$ since $\tilde\delta$ denotes a
smeared delta function. The smeared level density $\tilde g(E)$ (per
spin and isospin in this paper) in the denominator of expression
(\ref{eq3}) ensures the right normalization of $\hat{\rho}_E$. The
smooth quantities entering in (\ref{eq3}) are evaluated by replacing
$\hat{H}$, the independent-particle Hamiltonian, by its classical
counterpart $H_{cl}$ which corresponds to the Thomas-Fermi (TF)
approximation \cite{Ring,Brack}. This approach is not limited to the
evaluation of expectation values of single particle operators. Also
the average behavior of two-body matrix elements can be calculated
\cite{VSFC}. In this paper we are interested in the semiclassical
evaluation of the average pairing matrix elements which at TF level
read
\begin{equation}
v(E,E') = \frac{1}{\tilde g(E) \tilde g(E')} \sum_{\nu,\nu'} \tilde
\delta (E - \varepsilon_\nu) \, \tilde \delta(E' - \varepsilon_{\nu'}) \,
{\langle \Phi(\nu, \bar{\nu}) | v | \Phi(\nu', \bar{\nu}') \rangle} ,
\label{eq4} \end{equation}
where $|\Phi(\nu, \bar{\nu})\rangle$ is an {\em antisymmetric}
normalized two-body state constructed out of a state $| \nu \rangle$
and its time-reversed state $| \bar{\nu} \rangle$.
As it is
known \cite{Ring,Brack,R01}, the Strutinsky method averages the density matrix
over an energy interval corresponding roughly to the distance between
two major shells. Implicitly the same holds if the equivalent
Wigner-Kirkwood expansion (TF approximation at lowest order) is used
for obtaining $\hat{\rho}_E$.

As far as we are interested in the semiclassical counterpart of the
density matrix $\hat{\rho}_E$ on the energy shell, we start
considering its Wigner transform $f_E(\br,\bp)$. In order to obtain
the pure TF approximation, we differentiate with respect to $E$ the
Wigner-Kirkwood expansion of the full single-particle one-body density
matrix $\hat{\rho} = \Theta (E - \hat{H})$ retaining only the leading
term, which reads after normalization:
\begin{equation}
f^{TF}_E(\br,\bp) = \frac{1}{\tilde g(E)}\delta(E -H_{cl}),
\label {eq5} \end{equation}
where $H_{cl}= p^2/2{m^*}+V(\br)$ is the classical Hamiltonian of
independent particles with a constant effective mass $m^*$ moving in
an external potential well. Integration over the momentum yields the
local density on the energy shell:
\begin{equation}
\rho^{TF}_E(\br) =
\frac{1}{(2 \pi \hbar)^3} \int
d^3p f^{TF}_E(\br,\bp) =
\frac{m^* k_E(\br)}{2 \pi^2 \hbar^2 \tilde
g(E)},
\label{eq6} \end{equation}
where the local momentum at the energy $E$ is
\begin{equation}
k_E(\br) = \frac{p_E(\br)}{\hbar} = \sqrt{\frac{2m^*}{\hbar^2}(E - V(\br))}
\end{equation}
and the level density $\tilde{g}(E)$ is given by the integral of the local
level density $\tilde{g}(E,\br)$
\begin{equation}
\tilde{g}(E) = \int d^3 r \tilde{g}(E,\br)
= \int d^3r \frac{m^*k_E(\br)}{2\pi^2 \hbar^2}
\label{eq7a} \end{equation}
Now we proceed to calculate the average pairing matrix elements
$v(E,E')$ of the Gogny D1S force \cite{R5}
which is known to reproduce the experimental gap values when used in
microscopic Hartree-Fock-Bogolyubov calculations \cite{R51}.
Starting from (\ref{eq4}) and following the method explained in detail
in Ref. \cite{VSFC}, one arrives in TF approximation at
\begin{equation}
v(E,E') = \int d^3r \int \frac{d^3 p d^3p'}{(2 \pi \hbar)^6}
f^{TF}_{E}(\br,\bp) v(\bp - \bp')
f^{TF}_{E'}(\br,\bp')
\label{eq8} \end{equation}
where $f^{TF}_{E}$ is given by equation (\ref{eq5}) and $v(\bp-\bp')$
is the Fourier transform of the particle-particle part of the Gogny
force which describes the pairing. As far as the only dependence on
$\bp$ and $\bp'$ is in $v(\bp-\bp')$, we can average over the angle
between $\bp$ and $\bp'$ as follows:
\begin{equation}
v(p,p') =\frac{1}{4 \pi} \int v(\bp - \bp')
d \Omega.
\label{eq8a} \end{equation}
This result and the fact that the TF on shell density (\ref{eq5}) can
be recast as $f^{TF}_{E}(\br,\bp) = m^* \delta(p -p_E)/(\tilde{g}(E)
p_E)$ allows one to perform easily the angular integral in
(\ref{eq8}), with the result
\begin{equation}
v(E,E') = \frac{1}{\tilde{g}(E)\tilde{g}(E')} \frac{1}{4 \pi^3}
\bigg(\frac{2m^*}{\hbar^2} \bigg)^2
\int dr r^2 k_E k_{E'} v(p,p')
\label{eq8c} \end{equation}
which in the particular case of the Gogny force reads:
\begin{equation}
v(E,E') =  \sum_{i=1}^2  \frac{z_i}{\mu_i^2}
\frac{1}{2 \pi^3 \tilde{g}(E) \tilde{g}(E')} \bigg(\frac{2m^*}{\hbar^2}
\bigg)^2
\int_0^{R_t} dr r^2
\exp\big\{- \frac{\mu_i (k^2_E + k^2_{E'})}{4}\big\}
\sinh{\frac{\mu^2_i k_E k_{E'}}{2}} ,
\label{eq23} \end{equation}
where $R_t$ is the classical turning point and
$z_i = \pi^{3/2} \mu_i^3 (W_i - B_i - H_i + M_i)$.
The factors $z_i$ correspond to pairing in the $S=0$ and $T=1$
channel and are written in terms of the parameters of the Gogny
force $W_i$, $B_i$, $H_i$, $M_c$ and $\mu_i$ \cite{R5}.

The semiclassical TF gap equation reads
\begin{equation}
\tilde{\Delta}(E) = - \int^{V_2}_{V_1} d E' \tilde{v}(E,E') \tilde{g}(E')
\frac{\tilde{\Delta}(E')}{2 \sqrt{(E'- \mu)^2 + \tilde{\Delta}(E')^2}}
\label{eq9} \end{equation}
where $V_1$ and $V_2$ are the lower and upper limits of the pairing window
and $\mu$ is the chemical potential which is obtained by the condition of the
neutron (proton) number, i.e. integrating the corresponding TF level density
up to the Fermi level:
\begin{equation}
N_{\tau} = 2 \int_{V_0}^{E_{F_{\tau}}} \tilde{g}(E) dE,
\label{eq10} \end{equation}
where $V_0$ is the bottom of the potential
$V(\mbox{\boldmath$R$})$ and
$E_{F_{\tau}}$ is the Fermi level for each type of
particles ($\tau=$n,p).

The pairing density in the TF approximation is given by:
\begin{equation}
\tilde{\kappa}(E)= \frac{\tilde{\Delta}(E)}{2 \sqrt{(E - \mu)^2 +
\tilde{\Delta}(E)^2}}
\label{eq11} \end{equation}

Next we obtain the gap and the pairing density in coordinate space by
integrating over $E$ the gap and pairing density given by
Eqs. (\ref{eq9}) and (\ref{eq11}), repectively,
weighted with the local level density
$\tilde{g}(E,\mbox{\boldmath$R$})$:
\begin{equation}
\tilde{\Delta}(\br)= \int dE
\tilde{g}(E,\br) \tilde{\Delta}(E)
\label{eq12} \end{equation}
and
\begin{equation}
\tilde{\kappa}(\br) =  \int dE
\tilde{g}(E,\br)
\tilde{\kappa}(E),
\label{eq13} \end{equation}

In the calculation of the energy and $B(M1)$ factors of the scissors
mode the zeroth and fourth order moments of the pairing density for
each kind of nucleons are needed, they read:
\begin{equation}
k_0 = \int d^3r \tilde{\kappa}(\br)
= 4 \int dE \tilde{\kappa}(E)\int d^3r \tilde{g}(E,\br)
\label{eq14} \end{equation}
and
\begin{equation}
k_4 = \int d^3r \tilde{\kappa}(\br)x^2 z^2
= 4 \int dE \tilde{\kappa}(E)\int d^3r x^2z^2 \tilde{g}(E,\br),
\label{eq15} \end{equation}
where the factor four takes into account the
spin-isospin degeneracy.
In order to calculate Eqs.~(\ref{eq14}) and (\ref{eq15}),
we use a single particle potential of
harmonic oscillator type. The calculation of $\tilde{\kappa}(E)$
(Eq.~(\ref{eq11})) is carried out in spherical symmetry  and the
deformation is included in $\tilde{g}(E,\mbox{\boldmath$R$})$ in order to
obtain the moments of the pairing density. With the harmonic oscillator
potential the integral in coordinate space can be done analytically
and the calculation of the $k_0$ and $k_4$ moments reduce to the following
integrals over $E$:
\begin{equation}
k_0 = \frac{1}{2 \hbar^3 \omega_x^2 \omega_z}
\int dE\, E^2 \kappa(E)
\label{eq16} \end{equation}
and
\begin{equation}
k_4 = \frac{1}{6} \bigg(\frac{\hbar^2}{2 m^*}\bigg)^2
\frac{1}{\hbar^7 \omega_x^4 \omega_z^3}
\int dE\,E^4 \tilde{\kappa}(E),
\label{eq17} \end{equation}
where
$\omega_x^2$ and $\omega_z^2$ are given by formula (\ref{omegaxz})
and the deformation $\delta$ is
taken from the experiment \cite{Pietr}. In all the calculation an
effective mass $m^*=0.8m$ has been used.

\section*{Appendix D: Pair correlations in the superfluid model
of deformed atomic nuclei}

The microscopical calculations of integrals $k_0^\tau$ and $k_4^\tau$
(\ref{k0k4})
are performed with the single particle wave functions of the deformed
nuclei. These functions are obtained by the numerical solution of the
Schr$\ddot{\rm o}$dinger equation with the axially deformed Woods--Saxon
potential including the spin--orbit interaction:
$$
V_{WS}(\br)=-V_0/\{1+\exp[\alpha(r-R(\theta))]\},
$$
$$R(\theta)=R_0[1+\beta_0+\beta_2Y_{20}(\theta)+\beta_4Y_{40}(\theta)],
$$
$$V_{ls}(\br)=-\kappa(\bp\times \sigma)\nabla V_{WS}.$$
Here $R_0=r_0A^{1/3}$; the constant $\beta_0$ is introduced to ensure
the volume conservation;
$\beta_2$ and $\beta_4$ are quadrupole and hexadecapole deformation
parameters. The details of the method of the solution can be found in
\cite{Gareev0,Gareev1,Gareev2,Nester}.

The wave function of the deformed nucleus is represented in a form of
a superposition
$$
|\nu>\equiv \Psi_\Omega^\rho=\sum_{nlj} a_{nlj}^{\Omega
\rho}\cdot\Psi^\Omega_{nlj}, \eqno(63)
$$
where $a_{nlj}^{\Omega \rho}$ are the expansion coefficients and
$$
\Psi^\Omega_{nlj}=R_{nlj}(r){\bf Y}_{lj}^\Omega  \eqno(64)
$$
are the wave functions of the spherical basis.
Here  ${\bf Y}_{lj}^\Omega$ are the spherical spinors and $R_{nlj}(r)$ are
eigenfunctions of the radial part of the Schr$\ddot{\rm o}$dinger equation
with the spherical Woods--Saxon potential.
This paper deals with the single particle matrix elements of the type
 $<\nu|F(r)Y_{\lambda0|}\nu>$, with F(r) and $Y_{\lambda0}$ being the
radial part and the spherical function respectively.

Pair correlations are taken into account in the frame of the BCS theory.
The interaction leading to the superfluid pairing correlations acts
between the particles in time--reversed conjugate states. The pairing
matrix element $G(\nu+,\nu-;\nu'-,\nu'+)$ is usually assumed to
be a constant $G$ independent of $\nu$ and $\nu'$ \cite{Solov}. In this
approximation the gap $\Delta$ does not depend on $\nu$ either:
$\Delta=G\sum_\nu u_\nu v_\nu$. Then the equations determining
$\Delta$ and the chemical potential $\mu$ read:
$$
\frac{2}{G}=\sum_\nu \frac{ 1}{\epsilon_{\nu}},\eqno(66)\\
$$
$$
 N=2\sum_\nu v^2_\nu.
$$
Here N is the number of particles, $u_\nu$ and $v_\nu$ are the
coefficients of the Bogoliubov canonical transformation,
$\epsilon_\nu=(\Delta^2 +(E_\nu -\mu)^2)^{1/2}$ is the quasiparticle
energy, $E_\nu$ is the energy of the single particle state (63).
\begin{table}
\caption{\small  $\beta_2$, $\beta_4$, $V_0$ (in MeV), $r_0$ (in $fm$) and
$\alpha$ (in $fm^{-1}$) are parameters of the deformed Woods--Saxon
potential; $G$ (in MeV) are pairing strength constants; $\kappa$
are spin--orbit parameters in $fm^2$.}
 \label{table2}
\vspace{-0.2cm}
\begin{center}
\begin{tabular}{|c|c|c||c|c|c|c|c||c|c|c|c|c|}
\hline
   & & &
\multicolumn{5}{|c||} {Neutron system}    &
\multicolumn{5}{|c|} {Proton system}    \\
\cline{4-13}
  $A$ & $\beta_2$ & $\beta_4$ & $V_0$ & $r_0$ & $\alpha$ &$\kappa$ & $G$ & $V_0$ & $r_0$ & $\alpha$ &
 $\kappa$ & $G$  \\
\hline
 135 & 0.14 & 0    & 47.0 & 1.26 & 1.67 & 0.40 & 0.141 & 58.0 & 1.24 & 1.578 & 0.35 & 0.150 \\[-2mm]
 147 & 0.13 & 0    & 46.8 & 1.26 & 1.67 & 0.40 & 0.123 & 57.4 & 1.24 & 1.578 & 0.35 & 0.138 \\[-2mm]
 155 & 0.30 & 0.04 & 47.2 & 1.26 & 1.67 & 0.40 & 0.115 & 59.2 & 1.24 & 1.69  & 0.36 & 0.154 \\[-2mm]
 165 & 0.28 & 0.02 & 44.8 & 1.26 & 1.67 & 0.43 & 0.110 & 59.2 & 1.25 & 1.63  & 0.355& 0.132 \\[-2mm]
 173 & 0.26 &-0.02 & 44.8 & 1.26 & 1.67 & 0.42 & 0.108 & 59.2 & 1.25 & 1.59  & 0.32 & 0.133 \\[-2mm]
 181 & 0.20 &-0.03 & 43.4 & 1.26 & 1.67 & 0.40 & 0.106 & 59.8 & 1.24 & 1.67  & 0.33 & 0.130 \\[-2mm]
 193 & 0.14 & 0    & 43.4 & 1.26 & 1.67 & 0.40 & 0.101 & 59.8 & 1.24 & 1.67  & 0.33 & 0.121 \\
\hline
\end{tabular}
\end{center}
\end{table}
After all transformations the integrals $k_0^\tau$ and $k_4^\tau$ can
be written as
$$
k_0^\tau= 4 \int d^3 r \kappa_+^{\tau 0}(\br)=
4\sum_\nu u_\nu v_\nu \sum_{nlj} (a_{nlj}^{\nu})^2,\eqno(67)\\
$$
$$
k_4^\tau=4 \int d^3r x^2z^2\kappa_+^{\tau0}(\br)=\\
$$
$$
=\frac{8\sqrt{\pi}}{15}\sum_\nu u_\nu v_\nu<\nu|
r^4(Y_{00}+\frac{\sqrt{5}}{7}Y_{20}-\frac{4}{7}Y_{40})\kappa_+^{\tau
0}(\br)|\nu>=\\
$$
$$
= \sum_{\Omega,\rho} u_{\Omega\rho} v_{\Omega\rho}\sum_{nljn'l'j'}
a_{nlj}^{\Omega\rho} a_{n'l'j'}^{\Omega\rho}
I^{nlj}_{n'l'j'}\sum_{\lambda=0,2,4} L_\lambda C_{j\Omega,\lambda
0}^{j'\Omega} C_{j'1/2,\lambda 0}^{j1/2},\eqno(68)
$$
where $I^{nlj}_{n'l'j'}=\int r^6 R_{nlj}(r)R_{n'l'j'}(r)dr$,
coefficients $L_\lambda$ are: $L_0=\frac{4}{15}$,
$L_2=\frac{4}{15}$, $L_4=-\frac{16}{35}$,$ \\C_{j_1m_1,j_2m_2}^{JM}$
is the Clebsh--Gordan coefficient.

All considered nuclei were divided into several groups. The deformation
of the potential $V_{WS}$ in each group was chosen close to the average
value of experimental deformations of nuclei of the group. The schemes
of the single
particle states for each of these groups were calculated with the fixed
set of parameters, each set being fitted to achieve a correct sequence
of the single particle levels in deformed nuclei. The sets of these
parameters for neutron and proton systems are given in Table 2. The
parameters for A=155, 165, 173, 181 are taken from Ref. \cite{Malov}.
All discrete and quasidiscrete levels in the interval from the bottom
of the potential well up to the energy 22 MeV were taken into account
in the calculations. Such basis allows one to study low lying
states and giant resonances as well. The pairing strength constants $G$
were adjusted as a function of the size of the single particle basis.


\begin{thebibliography}{99}

\bibitem{BaSc2}
 E. B. Balbutsev and P. Schuck, Nucl. Phys. A {\bf 720} 293 (2003);\\
 E. B. Balbutsev and P. Schuck, Nucl. Phys. A {\bf 728} 471 (2003).

\bibitem{Zaw}
D. Zawischa, J. Phys. G: Nucl. Part. Phys. {\bf 24} 683 (1998).

\bibitem{Hilt92}
R. R. Hilton, Ann. Phys. (N.Y.)  {\bf 214} 258(1992).

\bibitem{Lo2000}
N. Lo Iudice, La Rivista del Nuovo Cimento {\bf 23} N9 (2000).

\bibitem{Bohle}
D. Bohle et al., Phys. Lett. B {\bf 137} 27 (1984).

\bibitem{Ziegler}
 W. Ziegler, C. Rangacharyulu, A. Richter, C. Spieler,
Phys. Rev. Lett. {\bf 65} 2515 (1990).

\bibitem{Pitz}
 H. H. Pitz et al., Nucl. Phys. A {\bf 509} 587 (1990).

\bibitem{Margraf}
J. Margraf et al., Phys. Rev. C {\bf 47} 1474 (1993).

\bibitem{Ranga}
 C. Rangacharyulu et al., Phys. Rev. C {\bf 43} R949 (1991).

\bibitem{Pietr1}
N. Pietralla et al., Phys. Rev. C {\bf 52} R2317 (1995).

\bibitem{Enders}
J. Enders et al., Phys. Rev. C {\bf 59} R1851 (1999).

\bibitem{Lipp}
E. Lipparini, S. Stringari, Phys. Lett. B {\bf 130} 139 (1983).

\bibitem{Strin}
E. Lipparini, S. Stringari, Phys. Rep. {\bf 175} 103 (1989).

\bibitem{Suzuki}
 T. Suzuki, D. J. Rowe, Nucl. Phys. A {\bf 289} 461 (1977).

\bibitem{Bes}
D. R. Bes, R. A. Broglia, Phys. Lett. B {\bf 137} 141 (1984).

\bibitem{Magnus}
I. Hamamoto, C. Magnusson, Phys. Lett. B {\bf 260} 6 (1991).

\bibitem{Pietr}
N. Pietralla et al., Phys. Rev. C {\bf 58} 184(1998).

\bibitem{Garrido}
E. Garrido et al., Phys. Rev. C {\bf 44} R1250 (1991).

\bibitem{Solov} V. G. Soloviev,
{\it Theory of Complex Nuclei} (Nauka, Moscow, 1971; Oxford, Pergamon Press, 1976).

\bibitem{Ring} P. Ring and P. Schuck,
 {\it The Nuclear Many-Body Problem} (Springer, Berlin, 1980).

\bibitem{Chand}
 S. Chandrasekhar, {\it Ellipsoidal Figures of Equilibrium}
 (Yale University Press, New Haven, Conn., 1969).

\bibitem{Bal}
 E. B. Balbutsev, Sov. J. Part. Nucl. {\bf 22} 159 (1991).

\bibitem{BaSc}
 E. B. Balbutsev and P. Schuck, Nucl. Phys. A {\bf 652} 221 (1999).

\bibitem{BM} A. Bohr and B. Mottelson,
 {\it Nuclear Structure} (Benjamin, New York, 1975), Vol. 2.

\bibitem{Hamam}
 I. Hamamoto and W. Nazarewicz, Phys. Lett. B {\bf 297} 25 (1992).

\bibitem{Matsuo}
 M. Matsuo, Y. Serizawa and K. Mizuyama, arXiv:nucl-th/0608048 v1.

\bibitem{Sushkov}
 V. G. Soloviev, A. V. Sushkov, N. Yu. Shirikova and N. Lo Iudice,
 Nucl. Phys. A {\bf 600} 155 (1996),\\
 V. G. Soloviev, A. V. Sushkov, N. Yu. Shirikova,
 Phys. Rev. C {\bf 53} 1022 (1996).

\bibitem{Macfar}
 M. Macfarlane, J. Speth and D. Zawischa,
 Nucl. Phys. A {\bf 606} 41 (1996).

\bibitem{Hilt86}
R. R. Hilton {\it et al.}, in {\it Proceedings of 1st International
Spring Seminar on Nuclear
Physics ``Microscopic Approaches to Nucler Structure", Sorrento, 1986},
Ed. by A. Covello (Bologna, Physical Society, 1986).

\bibitem{BaSc3}
 E. B. Balbutsev and P. Schuck, Physics of Atomic Nuclei {\bf 69} 1985 (2006).

\bibitem{VSFC}
X. Vi\~nas, P. Schuck, M. Farine and M. Centelles,
Phys.\ Rev. \ C {\bf 67} 054307 (2003).

\bibitem{Brack} M. Brack and R. K. Badhuri, {\it Semiclassical Physics}
(Addison-Wesley, Reading, MA, 1997).

\bibitem{R01}
M. Brack and H. C. Pauli, Nucl.\ Phys.\ A {\bf 207}, 401 (1973);
B. K. Jennings, Nucl.\ Phys.\ A {\bf 207} 538 (1973).

\bibitem{R5}
 J. F. Berger, M. Girod, and D. Gogny, Comput.\ Phys.\ Commun.\ {\bf
63} 365 (1991); Nucl.\ Phys.\ A {\bf 502} 85c (1989).

\bibitem{R51}
M. Kleban, B. Nerlo-Pomorska, J. F. Berger, J. Decharg\'e, M. Girod,
and S. Hilaire, Phys.\ Rev.\ C {\bf 65} 024309 (2002).

\bibitem{Gareev0}
F. A. Gareev, S. P. Ivanova, B. N. Kalinkin,
Izvestiya AN SSSR, ser. fiz. {\bf 33} 1690 (1968).

\bibitem{Gareev1}
F. A. Gareev, S. P. Ivanova, L. A. Malov and V.G. Soloviev,
Nucl. Phys. A {\bf 171} 197 (1971).

\bibitem{Gareev2}
 F. A. Gareev, S. P. Ivanova, V. G. Soloviev, S. I. Fedotov,
Sov. J. Part. Nucl. {\bf 4} 357 (1973).

\bibitem{Nester}
L. A. Malov, V. O. Nesterenko, N. Yu. Shirikova, JINR report,
R4--83--811 (Dubna, 1983).

\bibitem{Malov}
 L. A. Malov, V. G. Soloviev, Sov. J. Part. Nucl. {\bf 11} 301 (1980).


\end{thebibliography}
\end{document}